\documentclass[]{aastex}
\usepackage{emulateapj5}
\usepackage{onecolfloat}
\usepackage{graphicx} 
\usepackage{fancyheadings} 
\usepackage{ulem}
\usepackage{rotating}
\usepackage{lscape}

\newcommand{\medsizn}{+0.03}
\newcommand{\medcrfe}{+0.17}
\newcommand{\avnife}{+0.07}
\newcommand{\mednife}{+0.095}
\newcommand{\secqthr}{3.20}
\newcommand{\nNi}{31}
\newcommand{\nZn}{19}
\newcommand{\nSi}{35}
\newcommand{\nFeT}{50}
\newcommand{\ncrznfe}{12}

\newcommand{\ZSi}{<Z>^{\rm Si}}

\newcommand{\ZFe}{<Z>^{\rm Fe}}

\newcommand{\XFe}[1]{<[{\rm #1/Fe}]>}
\newcommand{\feh}{[{\rm Fe/H}]}
\newcommand{\sih}{[{\rm Si/H}]}
\newcommand{\znh}{[{\rm Zn/H}]}

\newcommand{\lya}{Ly$\alpha$ }

\newcommand{\cm}[1]{\, {\rm cm^{#1}}}
\newcommand{\N}[1]{{N({\rm #1})}}
\newcommand{\F}[1]{{F({\rm #1})}}

\newcommand{\fxi}[1]{{\xi({\rm #1})}}
\newcommand{\e}[1]{{\epsilon({\rm #1})}}

\newcommand{\rAA}{{\AA \enskip}}
\newcommand{\sci}[1]{{\rm \; \times \; 10^{#1}}}

\newcommand{\perd}{\;\;\; .}
\newcommand{\cmma}{\;\;\; ,}

\newcommand{\mkms}{{\rm \; km\;s^{-1}}}

\begin{document}

\submitted{Accepted to the Astrophysical Journal: Oct 15, 2001}

\title{The UCSD HIRES/KeckI Damped \lya Abundance Database\altaffilmark{1}
II. The Implications}

\author{ JASON X. PROCHASKA\altaffilmark{2,3}}
\affil{The Observatories of the Carnegie Institute of Washington}
\affil{813 Santa Barbara St.; Pasadena, CA 91101}
\email{xavier@ociw.edu}
\and
\author{ARTHUR M. WOLFE\altaffilmark{2}}
\affil{Department of Physics, and Center for Astrophysics and Space Sciences}
\affil{University of California, San Diego; 
C--0424; La Jolla; CA 92093}
\email{awolfe@ucsd.edu}

\begin{abstract} 

We present a comprehensive analysis of the damped \lya abundance
database presented in the first paper of this series.  
This database provides a homogeneous set of abundance measurements for
many elements including Si, Cr, Ni, Zn, Fe, Al, S, Co, O, and Ar from
38 damped \lya systems with $z_{abs} > 1.5$.  
With little exception, these damped \lya systems 
exhibit very similar relative abundances. 
There is no significant correlation in X/Fe with [Fe/H] metallicity and
the dispersion in X/Fe is small at all metallicity.

%

We search the database for trends indicative
of dust depletion and in a few cases find strong evidence. 
Specifically, we identify
a correlation between [Si/Ti] and [Zn/Fe] 
which is unambiguous evidence for depletion.  
Following \cite{hou01}, we present [X/Si] abundances against 
[Si/H]~$+ \log \N{HI}$ and note trends of decreasing
X/Si with increasing [Si/H]~$+ \log \N{HI}$ which argue for dust depletion. 
Similarly, comparisons of [Si/Fe] and [Si/Cr] against
[Si/H] indicate significant depletion at [Si/H]~$> -1$ but suggest
essentially dust-free damped systems at [Si/H]~$<-1.5$~dex. 

We present a discussion on the nucleosynthetic
history of the damped \lya systems by focusing on abundance patterns which
are minimally affected by dust depletion. 
We find [Si/Fe]~$\to +0.25$~dex
as [Zn/Fe]~$\to 0$ and that the [Si/Fe] values exhibit a plateau of
$\approx +0.3$~dex at [Si/H]~$<-1.5$~dex.  Together these trends
indicate significant $\alpha$-enrichment in the damped \lya systems
at low metallicity, an interpretation further supported by the 
observed O/Fe, S/Fe and Ar/Fe ratios.
Comparing the relative abundances of the Fe-peak elements, we identify
important offsets from solar relative abundances for Cr, Ni, and Fe
which suggest variations in nucleosynthesis along the Fe-peak.  
Finally, the damped \lya systems exhibit a modest odd-even effect 
revealed by Si/Al, [Si/Al]~$\approx +0.4$~dex,
which is significantly smaller than values observed
in Galactic halo stars of comparable metallicity. 
These observations present strong evidence 
that the damped \lya systems and Galactic halo
had different enrichment histories. 

To assess the impact of dust obscuration, we present estimates of the 
dust-to-gas ratios for the damped \lya sightlines and crudely calculate
dust extinction corrections. 
The distribution of extinction corrections suggests the
effects of dust obscuration are minimal and that the population
of 'missing' damped systems has physical characteristics similar
to the observed sample.

We update our investigation on the chemical evolution of the early
universe in neutral gas.  The results are in good agreement with
our previous work, but we emphasize two differences:
(1) the unweighted and $\N{HI}$-weighted [Fe/H] mean metallicities
now have similar values at all epochs except $z>3.5$ where small
number statistics dominate the $\N{HI}$-weighted mean;
(2)  there is no evolution in the mean [Fe/H] metallicity from
$z = 1.7 \to 3.5$ but possibly a marked drop at higher redshift.  

We conclude with a general discussion on the physical nature of the
damped \lya systems.  We stress the uniformity of the damped \lya
chemical abundances which
indicates that the protogalaxies identified with damped
\lya systems have very similar enrichment histories, i.e.
a nearly constant relative contribution from Type~Ia and Type~II SN.
The damped \lya systems also show constant 
relative abundances within a given system which places
strict constraints on the mixing
timescales of the damped systems and may pose a great challenge to the
protogalactic clump scenarios favored by hierarchical galaxy formation.


\end{abstract}

\twocolumn[%
]

\altaffiltext{1}{http://kingpin.ucsd.edu/$\sim$hiresdla} 
\altaffiltext{2}{Visiting Astronomer, W.M. Keck Telescope.
The Keck Observatory is a joint facility of the University
of California and the California Institute of Technology.}
\altaffiltext{3}{Hubble Fellow}

\pagestyle{fancyplain}
\lhead[\fancyplain{}{\thepage}]{\fancyplain{}{PROCHASKA \& WOLFE}}
\rhead[\fancyplain{}{The UCSD HIRES/Keck~I Damped \lya Abundance Database II}]{\fancyplain{}{\thepage}}
\setlength{\headrulewidth=0pt}
\cfoot{}

\section{INTRODUCTION}
\label{sec-intro}

Astronomers have long used high resolution 
spectroscopy to examine the chemical abundances of stars in
the Milky Way.  Besides meteoritic sampling, 
these studies provide the most accurate assessment of absolute and relative
nucleosynthetic yields in our Galaxy.  In turn, the results
test existing theories of nucleosynthesis 
as well as inspire new processes \citep[e.g.][]{whe89,edv93,mcw95}.
A principal goal of modern efforts is to extend these studies to 
the most extreme metal-poor stars whose enrichment might arise from
a few or even a single supernova.
Similar chemical abundance studies have been pursued within the
Galaxy's interstellar medium (ISM).  Spurred by Lyman Spitzer's inspiration,
UV spectroscopy acquired from space telescopes has led to a detailed
examination of the chemical and physical properties of this gas. 
Many of the same elements studied in stars are analyzed in the ISM, but in
contrast, the ISM
abundance patterns reflect more on processes of dust formation 
and destruction than nucleosynthesis
\citep[e.g.][]{spz93,sav96}. 

With the advent of 10m class telescopes, scientists
now perform ISM observations of high $z$ galaxies 
in a manner analogous to studies of the Milky Way ISM, i.e.,
through an analysis of high resolution spectroscopy of distant quasars
whose sightlines penetrate protogalaxies
(e.g.\ Wolfe et al.\ 1994; Prochaska \& Wolfe 1996, hereafter PW96;
Lu et al.\ 1996, hereafter L96; Prochaska \& Wolfe 1999, hereafter PW99;
Pettini et al.\ 2000).  
The majority of these
quasar absorption line (QAL) studies have 
focused on the chemical abundances of the damped \lya systems:
neutral hydrogen gas layers with HI column density, 
$\N{HI} \geq 2 \sci{20} \cm{-2}$.  
These QAL systems are the dominant reservoir of neutral baryons at
every observable epoch and are routinely identified as
the progenitors of modern galaxies.
Indeed, this was the founding principle of the first damped \lya surveys
\citep{wol86} and is well supported by the correspondence of the
inferred baryonic mass density of damped systems at $z \sim 2$ and the
mass density in stars today \citep{lzwt95,wol95,storr00}.

One of the most exciting aspects of the damped \lya systems is the fact that
they provide a means for assessing the chemical abundances of many 
elements for galactic systems in the early stages of evolution.  High resolution
studies of metal-poor stars may offer a complementary view, but at present these 
studies are limited to the Milky Way and a few other nearby systems.
Therefore, interpretations associated with metal-poor stars are strictly
limited by the fact that one is not sensitive {\it a priori} to
a wide range of star formation histories;  
there is always the possibility that the Milky
Way represents a unique path in the nucleosynthetic history
of the universe.  
In contrast, the damped \lya systems presumably probe protogalaxies with
a range of characteristics (e.g.\ mass, morphology, age).  At some level,
this diversity is a hindrance because our very
limited knowledge of these properties.  Nevertheless, the damped \lya
systems provide a valuable laboratory for research on stellar nucleosynthesis.

Because the damped \lya systems exhibit sub-solar metallicities, including
several with [Fe/H]~$< -2$,
one expects their observed abundance patterns to arise from 
a non-solar nucleosynthetic pattern modified by the 
depletion of refractory elements onto dust grains.
Ideally, one aims to disentangle the two patterns in order to
gauge the implications of dust depletion 
\citep[e.g.\ obscuration bias;][]{fall93} and reveal the chemical enrichment
history of these first galaxies.
Unfortunately, the analysis is confused by the 
limited number of routinely observed elements (Fe, Si, Cr, Ni, Al, Zn)
and the coincidence between a standard Type~II SN pattern and
the warm halo ISM dust depletion pattern 
(L96; PW99; Prochaska et al.\ 2000, hereafter P00).
The challenge is to identify abundance ratios which lift this degeneracy.

In this paper we analyze recently published chemical abundances of a large
database of $z>1.5$ damped \lya systems \citep[][Paper I]{pro01a}.
All of the data were obtained with the HIRES spectrograph \citep{vogt94}
on the Keck~I telescope, were reduced with the MAKEE software package,
and analyzed with the same techniques (see Paper~I for complete details).
It is a large homogeneous dataset, ideal for comprehensive analysis.
In the following, we restrict ourselves to five main areas:
($\S$~\ref{sec:elem}) an element by element discussion of abundance
trends with Fe metallicity; 
($\S$~\ref{sec:dust-depl}) an investigation of dust depletion;
($\S$~\ref{sec:nucleo}) implications for nucleosynthesis in these protogalaxies;
($\S$~\ref{sec:dust-obsc}) evidence for/against dust obscuration; 
and 
($\S$~\ref{sec:chemev}) the chemical enrichment history inferred 
from damped \lya systems.  We conclude ($\S$~\ref{sec:discuss}) with
a general discussion of the damped \lya systems including speculations
on their physical nature.  Given the length of this manuscript, we 
encourage the reader to
focus on areas of primary interest and refer to the summary tables
throughout.


\section{ELEMENTAL ABUNDANCES}
\label{sec:elem}

In this section we present the most important
elemental abundances for our sample of damped \lya systems, stressing
the new results from this paper. 
Throughout the analysis
we adopt no ionization corrections, i.e., we assume that the column density
of element X is equal to the column density of the dominant ionization
state in neutral hydrogen gas: Fe$+$ for Fe, O$^0$ for O, H$^0$ for H.
This assumption is theoretically motivated (e.g. PW96; Viegas 1994),
but not well established empirically.  
Recent calculations indicate ionization corrections should be small
for the majority of elements considered here \citep{vladilo01}, but
future damped \lya studies need to address this point unambiguously
\citep{pro02}.

Following standard practice in stellar abundance studies,
we plot the abundance of each element with respect to
Fe relative to the solar ratio, [X/Fe]~$\equiv \log [\N{X}/\N{Fe}] -
\log [\N{X}/\N{Fe}]_\odot$, against [Fe/H].  In a stellar analysis, this 
comparison highlights trends which reflect nucleosynthetic
processes convolved with the star formation history.  For the
damped \lya systems, the analysis is sensitive to the combined effects of 
nucleosynthesis and dust depletion.  
Unless otherwise noted, we plot $1\sigma$ error bars\footnote{Note the
error adopted is the quadrature sum of the errors in logarithmic space} 
derived from the apparent
optical depth method \citep[AODM;][]{sav91} as listed in Paper~I.
For each element, we consider several simple statistics 
(e.g.\ a Pearson's test to investigate linear correlations)
of the observed sample excluding all upper and lower limits.
A summary of this section is given in 
Tables~\ref{tab:Sumelm} and \ref{tab:absum} 
($\S$~\ref{sec:absum}).

\subsection{Fe-Peak Elements -- Cr, Mn, Co, Fe, Ni, Zn}

We begin with the elements whose atomic numbers are near Fe,
the so-called Fe-peak elements.  We include Zn, 
although it is not a true Fe-peak element and is probably produced
through different nucleosynthetic processes \citep{hff96,umeda01}.

\subsubsection{Iron}
\label{subsec:Fe}

Almost all of our interpretations of dust depletion, dust obscuration,
and nucleosynthesis are drawn in part from Fe measurements and,
therefore, it is important to highlight the
uncertainties associated with measuring $\N{Fe}$.  The most
significant uncertainties at the time of this publication are in
the oscillator
strengths of the Fe~II $\lambda\lambda$1608,1611 transitions.  For 
Fe~II 1608, the $f$-value is taken from \cite{bergs96} who attribute
a $9\%$ error to their laboratory value which translates to $\approx 0.035$~dex
uncertainty.  For Fe~II 1611, we consider the most reliable value
to be the theoretical value from \cite{raassen98}.  Until now, most damped
\lya analyses have relied upon the $f_{1611}$
value presented by \cite{cardelli95}
from their empirical analysis of several Fe~II transitions
observed in the ISM.  The difference is
significant; the \cite{raassen98} $f$-value implies an $\approx 0.1$~dex lower
$\N{Fe^+}$ value than the \cite{cardelli95} value.  As discussed in Paper~I,
the $\N{Fe^+}$ value derived from Fe~II 1611 is in good agreement with the
values from Fe~II 2249 and 2260 in the systems where we observed all
three profiles.  Therefore, we are reasonably confident in the adopted 
Fe~II 1611 oscillator strength.

For some of the following analysis, we adopt a proxy for Fe (e.g. Cr, Ni, Al)
when we have no reliable $\N{Fe^+}$ measurement.
In these cases, we assume [Fe/H] = [Ni/H]~$- \, 0.1$, 
[Fe/H] = [Cr/H]~$- \, 0.2$,
or [Fe/H] = [Al/H] for reasons which will become obvious below.
When we have made this substitution, we plot these specific data points
with a unique symbol and generally
do not include their values in any statistical
analysis (i.e. means, medians, Pearsons' tests).

\subsubsection{Chromium}

\begin{figure}[ht]
\includegraphics[height=3.8in, width=2.8in,angle=-90]{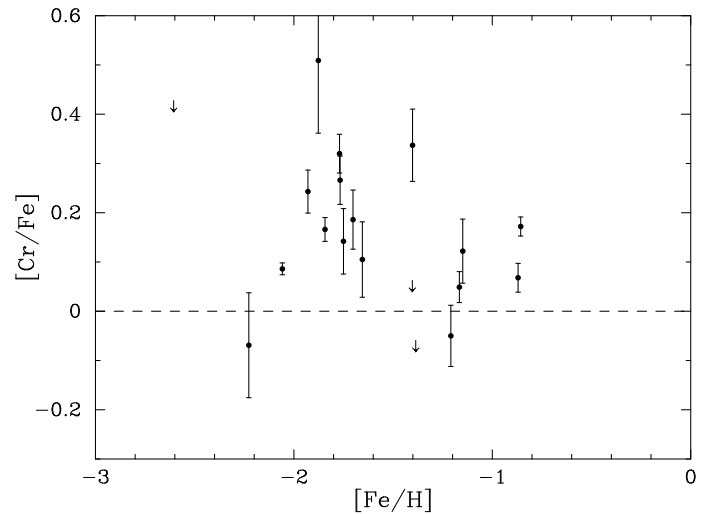}
\begin{center}
\caption{ [Cr/Fe] abundance ratios vs.\ [Fe/H] metallicity for our
complete sample of damped \lya systems.  Nearly every system shows
a mild enhancement and there is a mild trend of decreasing Cr/Fe
with increasing metallicity (although $P_{Pears} = 0.52$ due to 
the system at [Fe/H]~$= -2.2$).
The dashed line at [Cr/Fe] = 0 indicates 
the solar meteoritic Cr/Fe ratio.
}
\label{fig:CrFe}
\end{center}
\end{figure}

Figure~\ref{fig:CrFe} presents the [Cr/Fe] values vs.\ [Fe/H]
for our complete sample. Nearly every damped system shows
an enhanced Cr/Fe ratio,  a behavior first noted in L96 and
emphasized in PW99.  The median enhancement is \medcrfe~dex 
which is suggestive of dust
depletion because no stellar population shows enhanced Cr/Fe whereas
dust depleted ISM sightlines do exhibit a mild Cr overabundance.
One also notes a slight anti-correlation between [Cr/Fe] and metallicity.
If the trend is not the result of small number statistics,  
it may have significant implications for dust depletion 
($\S$~\ref{subsec:XFeZnFe}).



\begin{figure}[hb]
\includegraphics[height=3.8in, width=2.8in,angle=-90]{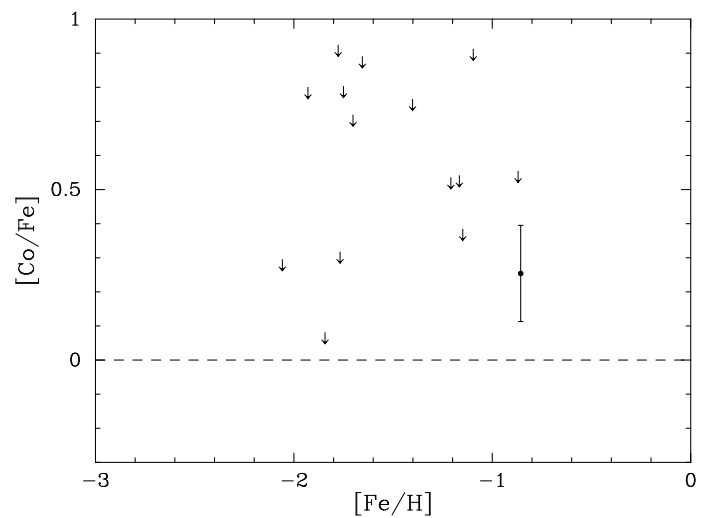}
\begin{center}
\caption{[Co/Fe] abundance ratios vs.\ [Fe/H] metallicity for our
complete sample of damped \lya systems.  
We report only a single detection and are even suspect of its value.
With improved S/N, several of these systems might present meaningful
constraints on Co/Fe.
The dashed line at [Co/Fe] = 0 indicates the solar meteoritic Co/Fe ratio.
}
\label{fig:CoFe}
\end{center}
\end{figure}

\subsubsection{Cobalt}

\cite{ellison01} recently drew attention to the prospect of measuring Co
in the damped \lya systems.  
Figure~\ref{fig:CoFe} plots the [Co/Fe] values for our full sample,
with all but one measurement an upper limit on $\N{Co^+}$.  
We report a measurement for the system at $z_{abs} =1.920$ 
toward Q2206--19
which is consistent with the value reported by \cite{ellison01}.
Our value is based on the very weak Co~II 1574 transition, however,
and is
therefore highly suspect.  Furthermore, the upper limits that we place
on $\N{Co^+}$ from Co~II 1941 and 2012 are smaller than the value for
Co~II 1574 and also are in conflict with the value implied by Ellison et al.
It is possible that the differences are due to continuum error, although
we suggest Ellison et al.\ slightly overestimated $\N{Co^+}$.  

The effect of dust depletion on the Co/Fe ratio is not well understood,
in particular because Co$^+$ has been very rarely detected in the ISM.
From the measurements for $\zeta$~Oph, it appears that Co has a 
refractory nature similar to Fe and the other Fe-peak elements.  
This is expected given the correlation observed between the gas-phase
abundance and condensation temperature of many elements 
\citep[e.g.][]{jenkins87}.  Therefore,
a significant departure from the solar ratio might indicate 
important nucleosynthetic processes \citep[e.g.][]{mcw97}.  
Figure~\ref{fig:CoFe} indicates several upper limits whose value approach
the solar abundance.
Greater attention should be given to these few systems where 
improved S/N can provide meaningful Co measurements.

\begin{figure}[ht]
\includegraphics[height=3.8in, width=2.8in,angle=-90]{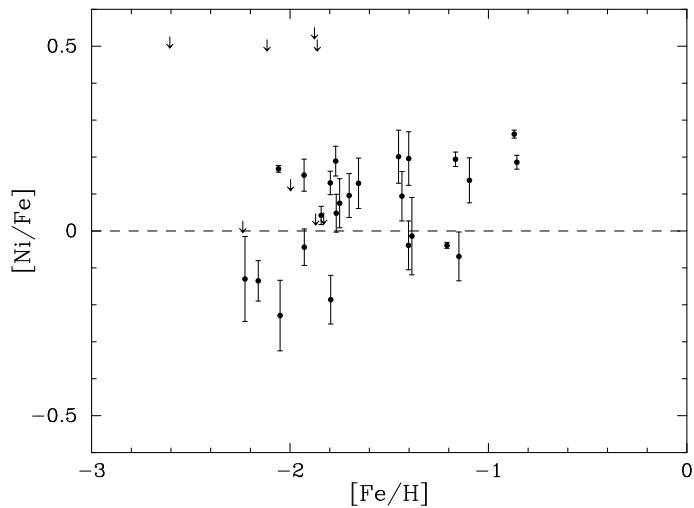}
\begin{center}
\caption{[Ni/Fe] abundance ratios vs.\ [Fe/H] metallicity for our
complete sample of damped \lya systems.  
The values scatter about the solar ratio with the majority exhibiting
a small enhancement.
We also note a significant trend with metallicity ($P_{Pears} = 0.02$).
The dashed line at [Ni/Fe] = 0 indicates the solar meteoritic Ni/Fe ratio.
}
\label{fig:NiFe}
\end{center}
\end{figure}

\subsubsection{Nickel}
\label{subsec:Ni}


Figure~\ref{fig:NiFe} plots [Ni/Fe] vs.\ [Fe/H] for \nNi\ of 
our damped \lya systems.
Although the [Ni/Fe] points scatter about the solar value, the
majority are mildly enhanced.  
The unweighted mean of all measurements (no limits) is 
$\XFe{Ni} \, = \avnife \pm 0.03$ and the median value is \mednife.  
This enhancement is not observed in the ISM and
conflicts with empirical trends in metal-poor stars although
a Ni/Fe enhancement is predicted in core-collapse explosions of 
massive Pop~III stars \citep{umeda01}.  
Of course, a small error may 
exist in the Ni$^+$ or Fe$^+$ atomic data implying this mild enhancement.
An enhanced
Ni/Fe ratio can also arise from photoionization \citep{howk99}, but one would
also predict a large Ni/Cr ratio where the opposite is observed.
Finally, it is possible the enhanced Ni/Fe values reflect intrinsic
differences in abundance patterns between stellar atmospheres and 
the damped \lya systems.

\subsubsection{Zinc}
\label{sec:Zn}

Zinc plays a pivotal role in interpreting the abundance patterns of the
damped \lya systems.  Since \cite{ptt94} demonstrated that most damped
systems exhibit enhanced Zn/Cr ratios, the community has generally 
accepted the notion that the refractory elements (e.g.\ Fe, Ni,	Cr)
are mildly depleted.  Because of the impact of this element, it is
important to carefully examine any issues which influence measurements
of $\N{Zn^+}$.  

While there are two strong Zn~II transitions at 2026\rAA
and 2062\AA,  the latter transition is often blended with the 
Cr~II 2062 transition.  Therefore we rely primarily on the Zn~II 2026 profile,
in particular because the AODM does not easily account for line blending.
The difficulty, however, is that the Zn~II 2026 profile may also 
be blended with
two other transitions: Mg~I 2026 and Cr~II 2026 with separations 
from Zn~II 2026 of
+52~km/s and +20~km/s respectively.  Luckily, the oscillator strength
of the Cr~II 2026 transition is small enough that it should
not significantly affect the $\N{Zn^+}$ measurement. 
The Mg~I 2026 profile, meanwhile, has been 
considered in previous works \citep[e.g.][]{ptt94}, but was generally
ignored in PW99.  The difficulty with this transition is that Mg$^0$
is not the dominant ionization state of Mg in a neutral hydrogen
gas region; it has an ionization potential IP~$< 13.6$~eV.  Therefore,
the population of this ionization state is sensitive to a variety of
physical conditions all of which are largely unconstrained by our
observations.  While another strong Mg~I transition does exist at 
2796\AA, our observations never cover this profile.  In short, those 
systems which have absorption profiles spanning over 50~km/s should be 
considered to have $\N{Zn^+}$ values with upper limits until Mg$^0$
is carefully considered.  

An excellent example of this issue is presented by 
the damped system at $z=1.776$ toward Q1331+17 (Paper~I, $\S$~\secqthr).  
\cite{ptt94} noted a significant feature in the 
Zn~II 2026 profile which coincides with the Mg~I 2026 transition and
corrected $\N{Zn^+}$ accordingly.  At higher resolution, however, we
find that this absorption also coincides with a significant feature
in the Fe~II profiles meaning it could be another component
in the Zn~II profile, not Mg~I.  Even more confusing is the fact that the
$\N{Zn^+}$ values for Zn~II 2026 and 2062 agreed very well in our
initial analysis (PW99).
After performing a line-profile fitting analysis
of all of the Cr~II and Zn~II profiles, however, we
derived an $\N{Zn^+}$ value $\approx 0.05$~dex lower than our previous 
analysis and a much lower $\N{Mg^0}$ value than that inferred by
Pettini et al.  
In this case, therefore, correcting for Mg$^0$ suggests
a significant underestimate of Zn$^+$ in the moderate resolution
analysis but implies our initial echelle analysis provided
a small overestimate to $\N{Zn^+}$.

\begin{figure}[ht]
\includegraphics[height=4.8in, width=3.8in]{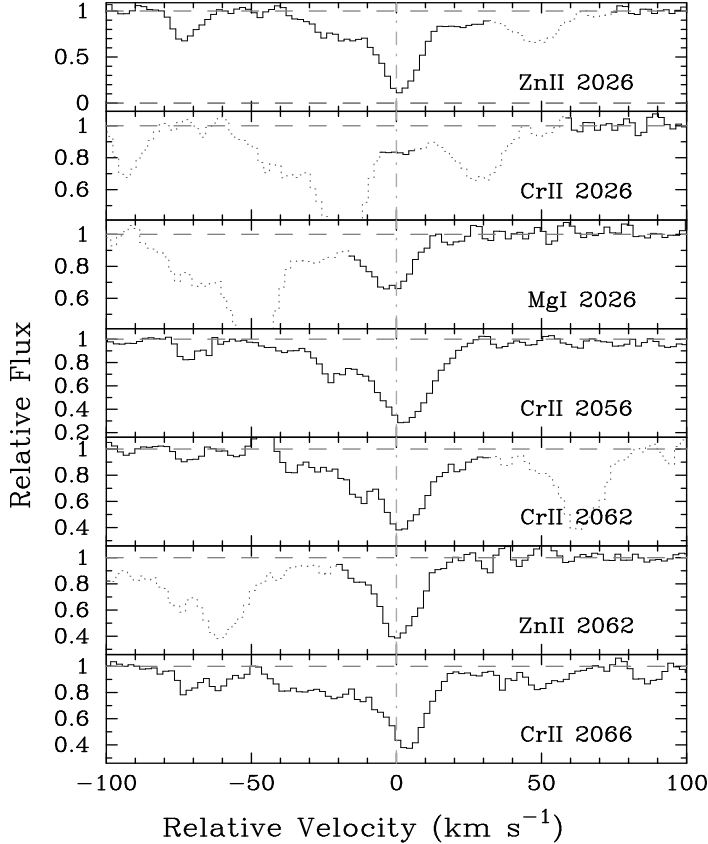}
\begin{center}
\caption{Metal-line profiles of the Zn~II and Cr~II transitions
near 2025\AA\ and the Mg~I 2026 transition for the damped \lya system
toward Q0458--02.
A comparison of the profiles
provides convincing evidence that the feature at $v = 50 \mkms$ in
the Zn~II 2026 profile is not Zn$^+$ absorption but is due to 
Mg~I 2026.  Similar blending must be considered when measuring $\N{Zn^+}$
from the Zn~II 2026 transition.
It is also possible, although far less certain, that
we identify absorption due to Cr~II 2026.
}
\label{fig:q0458}
\end{center}
\end{figure}

An unequivocal example of Mg~I 2026 absorption in our sample is
in the damped \lya system at $z_{abs}=2.040$ toward Q0458$-$02.  
Figure~\ref{fig:q0458} presents the Zn~II 2026, Mg~I 2026, and Cr~II 2026
profiles as well as the Cr~II 2056, Cr~II 2062, and Zn~II 2062
transitions for this system.  It is clear that the strong component
at $v=0$~km/s in the Mg~I profile must be attributed to Mg$^0$ and not Zn$^+$.  
If we assume that the Mg~I profile tracks the low-ion profiles (an 
assumption which need not hold), then the Mg~I profile at $v<50 \mkms$
contributes
$\approx$4 m\rAA to the Zn~II 2026 equivalent width.  This would imply
a 0.025~dex enhancement to $\N{Zn^+}$ in the linear limit of the curve
of growth but more likely corresponds to a $\approx$0.020~dex correction.
In Paper~I we revised our $\N{Zn^+}$ measurement 
downward by 0.02~dex from the value reported in PW99 to account for
this blending. 
In Table~\ref{tab:zncrmg}, we list all of the systems in our analysis
with Zn$^+$ measurements, note the likelihood of Mg~I/Zn~II line-blending,
estimate a correction to $\N{Zn^+}$, and indicate whether this 
correction was applied. 
In general, unless we identified
clear evidence for the presence of Mg~I, we chose not to apply any
correction to $\N{Zn^+}$.  In the future, we will update the values in
our online database as measurements of the Mg~I 2796 profile are obtained.


\begin{table}[ht]
\begin{center}
\caption{
{\sc Zn MEASUREMENTS \label{tab:zncrmg}}}
\begin{tabular}{lccccc}
\tableline
\tableline
QSO &$z_{abs}$ & ZnII/MgI & Estimated & $\N{Zn^+}$ \\
& & Blending? & correction & corrected? \\
\tableline
PH957       & 2.309 & no  & \\
Q0149+33    & 2.141 & no  & ... \\
Q0201+36    & 2.463 & ?   & ...  & no \\
Q0458--02   & 2.040 & yes & 0.02 & yes \\
Q0841+12    & 2.375 & no  & ... \\
Q0841+12    & 2.476 & no  & ... \\
Q1210+17    & 1.892 & no  & ... \\
Q1215+33    & 1.999 & no  & ... \\
Q1223+17    & 2.466 & no  & ... \\
Q1331+17    & 1.776 & yes & 0.06 & yes \\
Q2206--19   & 1.920 & yes & ?    & no \\
Q2230+02    & 1.864 & yes & ?    & no \\
Q2359--02   & 2.095 & ?   & ?    & no \\
\tableline
\end{tabular}
\end{center}
\end{table}
 
\begin{figure}[ht]
\includegraphics[height=3.8in, width=2.8in,angle=-90]{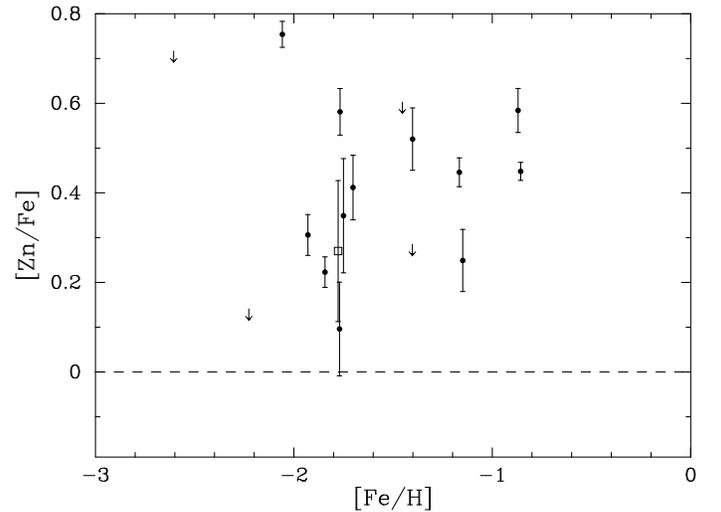}
\begin{center}
\caption{[Zn/Fe] abundance ratios vs.\ [Fe/H] metallicity for our
complete sample of damped \lya systems.  
Every system shows an enhanced value and there is no trend with 
metallicity.
The dashed line at [Zn/Fe] = 0 indicates the solar meteoritic Zn/Fe ratio.
The values plotted with a black square indicate that a proxy was substituted
for Fe and an additional 0.05~dex error is being assumed.
}
\label{fig:ZnFe}
\end{center}
\end{figure}

Figure~\ref{fig:ZnFe} presents the 
[Zn/Fe] values vs.\ [Fe/H] for our complete 
sample.  Essentially every system exhibits an enhanced Zn/Fe ratio.   
While there are a number of systems with [Zn/Fe]~$\leq 0.3$~dex 
which could be explained by nucleosynthesis (e.g. P00),
there are several systems with [Zn/Fe]~$> 0.4$~dex which
are highly suggestive of dust depletion.  
Although it could be small number statistics, four of the six systems
with [Fe/H]~$> -1.5$~dex exhibit large Zn/Fe ratios. 
Despite this, there is no significant
linear correlation between [Zn/Fe] and [Fe/H] 
(but see $\S$~\ref{sec:SiH}).  The observed
range of Zn/Fe ratios may indicate a significant
range in dust-to-gas ratios in the damped \lya systems from system to system
or at least sightline to sightline as noted by \cite{ptt97}.  
We warn, however, that a fraction of this scatter 
could be related to the line blending
described above or stochastic Zn nucleosynthesis.

\begin{figure}[ht]
\includegraphics[height=3.8in, width=2.8in,angle=-90]{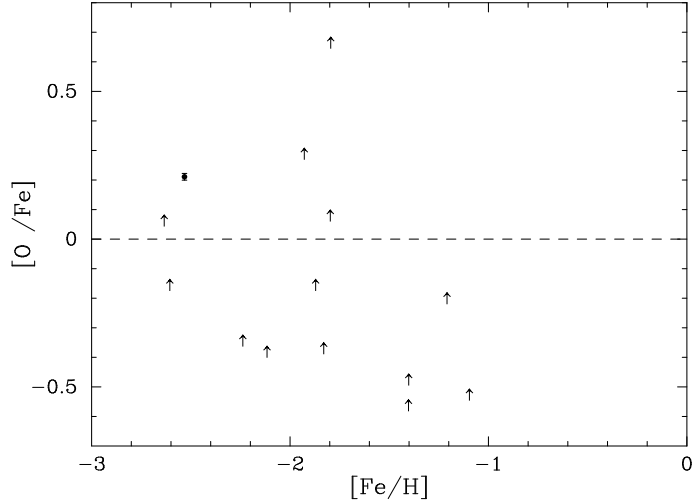}
\begin{center}
\caption{[O/Fe] abundance ratios vs.\ [Fe/H] metallicity for our
complete sample of damped \lya systems.  
With only one exception, all of the O~I profiles are saturated in 
our damped \lya systems and we only present lower limits on the O/Fe 
ratio.  Only in the lowest metallicity systems do our [O/Fe] values
begin to place meaningful constraints on the $\alpha$-enrichment
of the systems.  Future observations with higher UV sensitivity
should provide more measurements.
The dashed line at [O/Fe] = 0 indicates the solar photospheric Ar/Fe ratio.
}
\label{fig:OFe}
\end{center}
\end{figure}

\subsection{Alpha Elements}
\label{subsec:alpha}


\subsubsection{Oxygen}

Oxygen is the prototypical $\alpha$-element and its nucleosynthesis
is dominated by Type~II SN. Furthermore, it offers a particularly
promising avenue for probing nucleosynthesis in the damped \lya systems
because it is only mildly refractory.  Unfortunately, it is very
challenging to observe oxygen in the damped systems; the majority of lines
lie within the \lya forest and tend to be heavily saturated.
\cite{molaro00} have recently claimed a measurement of oxygen in 
the damped \lya system toward Q0000$-$26 based on an analysis of
the O~I $\lambda\lambda$ 925,950 transitions.  
Adopting the \cite{raassen98} oscillator strength for
Fe~II 1611, these observations indicate an enhancement
[O/Fe]~$\approx +0.3$~dex consistent with values derived in metal-poor stars.

Figure~\ref{fig:OFe} presents our sample of [O/Fe] measurements
against [Fe/H] metallicity.
With the exception of one system,
all of our measurements on oxygen are lower limits.
Unfortunately, only at the lowest
[Fe/H] values do these limits begin to place meaningful constraints on
the O/Fe ratio.  The single measurement comes from the system at 
$z_{abs} = 2.844$ toward Q1946+76.  L96 placed a lower limit on $\N{O^0}$
from the saturated O~I 1302 transition, but our extensive wavelength
coverage enables a measurement of $\N{O^0}$ from the O~I 948 and 1039 
transitions.  The observed ratio, [O/Fe]~$\approx +0.2$, indicates a modest
enhancement.  Surprisingly, the $\N{O^0}$ measurement also implies a 
sub-solar O/Si ratio which contradicts the observed trend in metal-poor
stars \citep[e.g.][]{mcw95}.  Perhaps this value is 
indicative of photoionization; Si$^+$ can arise in regions
of photoionized gas.
The other notable value is the lower limit on O/Fe 
for the system toward Q0336-01 ([O/Fe] = +0.6~dex which suggests an 
extremely $\alpha$-enriched system.

\begin{figure}[ht]
\includegraphics[height=3.8in, width=2.8in,angle=-90]{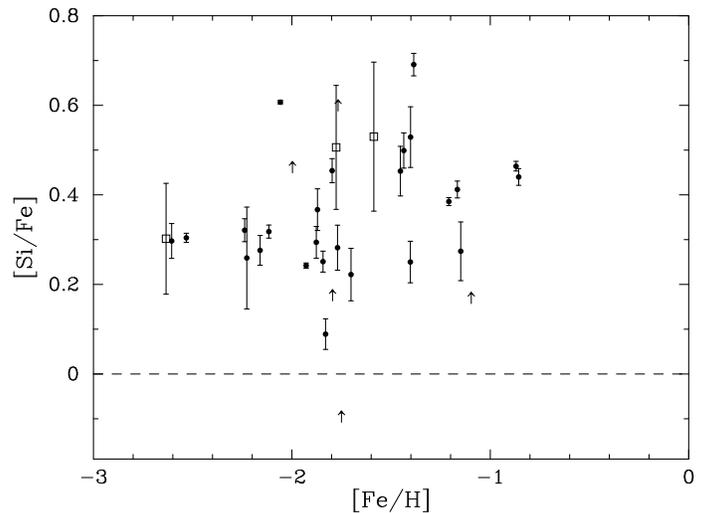}
\begin{center}
\caption{[Si/Fe] abundance ratios vs.\ [Fe/H] metallicity for our
complete sample of damped \lya systems.  
Every system shows an enhanced value and there is no trend with metallicity.
The dashed line at [Si/Fe] = 0 indicates the solar meteoritic Si/Fe ratio.
The values plotted with a black square indicate that a proxy was substituted
for Fe and an additional 0.05~dex error is being assumed.
}
\label{fig:SiFe}
\end{center}
\end{figure}

\subsubsection{Silicon}
\label{subsec:Si}

With few exceptions, silicon represents the only
$\alpha$-element measurement for a given damped \lya system.
Although silicon is a refractory element, its depletion is mild in 
lightly depleted regions of the
ISM and therefore the observed column densities for the damped \lya
systems should nearly reflect the total Si column density.
Figure~\ref{fig:SiFe} plots the Si/Fe ratios against [Fe/H] metallicity
for the \nSi\ systems with a Si measurement.  All of the damped \lya
systems show enhanced Si/Fe ratios with values ranging from +0.1 to +0.6~dex.
Examining Figure~\ref{fig:SiFe},
one notes the mean and scatter in the
Si/Fe ratios exhibit little variation with [Fe/H] except that the systems
at [Fe/H]~$< -2$~dex show nearly constant Si/Fe ratios.
The mean [Si/Fe] enhancement is similar to that for [Zn/Fe]
which has led several authors to argue for a nearly solar Si/Fe ratio
when dust depletion is taken into account \citep[e.g.][]{vladilo98}.
As discussed in greater detail in $\S$~\ref{sec:nucleo}, however,
we contend most systems exhibit a nucleosynthetic Si/Fe
enhancement suggestive of Type~II SN enrichment.

\subsubsection{Sulfur}

\begin{figure}[ht]
\includegraphics[height=3.8in, width=2.8in,angle=-90]{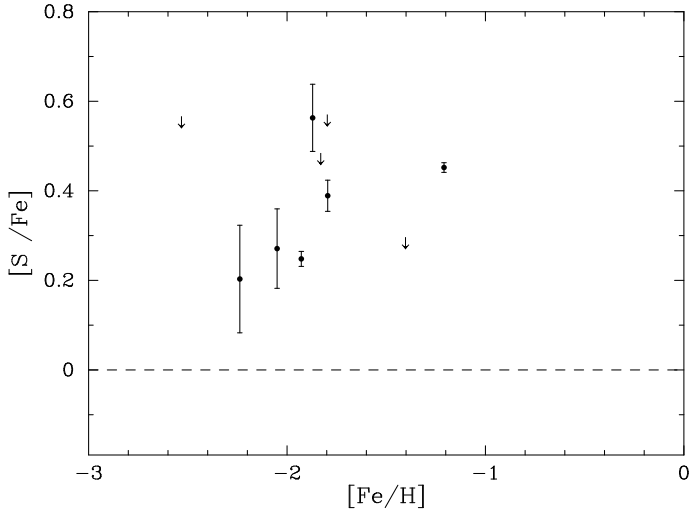}
\begin{center}
\caption{[S/Fe] abundance ratios vs.\ [Fe/H] metallicity for our
complete sample of damped \lya systems.  
Every system shows an enhanced value and there is no obvious
trend with metallicity.
The dashed line at [S/Fe] = 0 indicates the solar meteoritic S/Fe ratio.
}
\label{fig:SFe}
\end{center}
\end{figure}

Sulfur is one of the truly non-refractory elements in the ISM, therefore
it provides an unambiguous probe of nucleosynthesis.
Figure~\ref{fig:SFe} presents our complete sample of measurements. 
Like silicon, the S/Fe ratios are modestly enhanced in all cases.

\clearpage

\begin{figure}[ht]
\includegraphics[height=3.8in, width=2.8in,angle=-90]{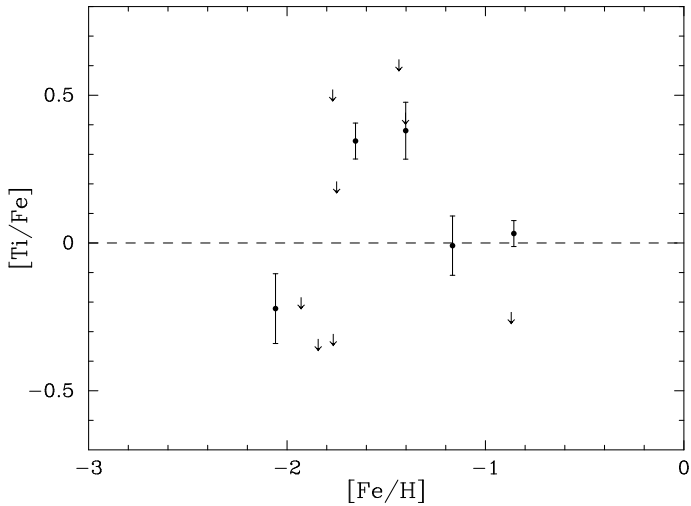}
\begin{center}
\caption{[Ti/Fe] abundance ratios vs.\ [Fe/H] metallicity for our
complete sample of damped \lya systems.  
The systems exhibit a range of values including several sub-solar
upper limits which are indicative of dust depletion.
The dashed line at [Ti/Fe] = 0 indicates the solar meteoritic Ti/Fe ratio.
}
\label{fig:TiFe}
\end{center}
\end{figure}

\subsubsection{Titanium}
\label{subsec:TiFe}

On theoretical grounds, Ti is expected to track Fe or even exhibit
sub-solar abundances in the Galaxy \citep[e.g.][]{timmes95}.
Empirically, however, Ti generally
tracks the other $\alpha$-elements with enhancements relative to Fe 
of $\approx +0.3$~dex in metal-poor stars \citep[][P00]{mcw95}.  
In the ISM, Ti either tracks Fe or is 
deficient relative to Fe \citep{sav92,howk99b}.  
Significant departure of Ti/Fe from the
solar ratio, therefore, should provide unambiguous evidence for
either nucleosynthesis or dust depletion.

Figure~\ref{fig:TiFe} presents our
new and revised [Ti/Fe] values against [Fe/H]. For the few systems with
measurements, the values scatter about the solar ratio.   
Having corrected for the oscillator strengths from PW99
\citep{wiese01}, we no longer
find significant evidence for an enhancement of Ti/Fe in the damped \lya
systems.  In fact, the most striking aspect of Figure~\ref{fig:TiFe} is
the set of three upper limits with [Ti/Fe]~$< -0.3$~dex
which suggest significant dust depletion.
In one case (Q1223+17), however, the observed Zn/Fe
ratio is small ([Zn/Fe]~$\approx +0.2$~dex) 
and the system is unlikely to be severely
depleted.  Although it is a single example,
one must be cautious in interpreting the low Ti/Fe ratios 
solely in terms of dust depletion. 

\begin{figure}[ht]
\includegraphics[height=3.8in, width=2.8in,angle=-90]{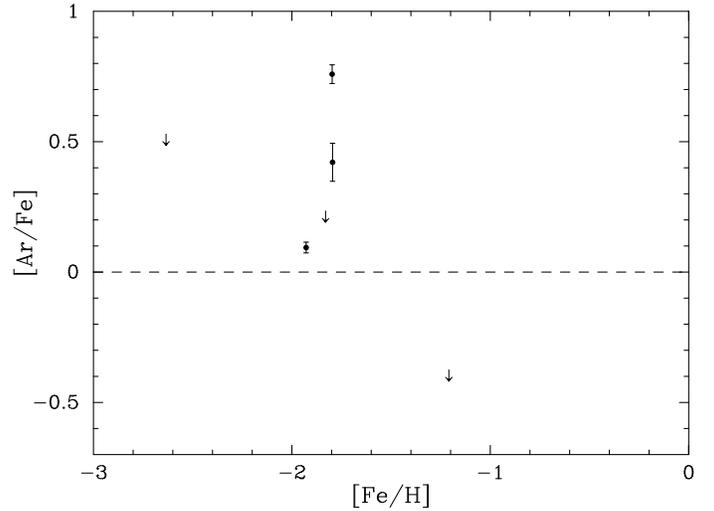}
\begin{center}
\caption{[Ar/Fe] abundance ratios vs.\ [Fe/H] metallicity for our
complete sample of damped \lya systems.  
The majority of systems show solar or enhanced Ar/Fe ratios suggestive
of Type~II SN enrichment and/or dust depletion.  
The dashed line at [Ar/Fe] = 0 indicates the solar photospheric Ar/Fe ratio.
}
\label{fig:ArFe}
\end{center}
\end{figure}

\subsubsection{Argon}

Although argon is presumed to track the other $\alpha$-elements,
there is little empirical evidence.
Argon is rarely observable in stellar atmospheres but it does
trace O and S in metal-poor H~II regions \citep{henry99}.
Unfortunately,
our observations provide measurements on Ar for only a limited number of 
systems.  Figure~\ref{fig:ArFe} plots these observations against 
[Fe/H] metallicity.  In agreement with the result presented by
\cite{molaro01} for Ar in the system toward Q0000--26,
several of our systems show enhanced Ar/Fe ratios.  
Because Ar is not depleted onto dust grains, the enhanced Ar/Fe values 
could reflect dust depletion, but they are also suggestive
of significant Type~II SN enrichment.

\begin{figure}[ht]
\includegraphics[height=3.8in, width=2.8in,angle=-90]{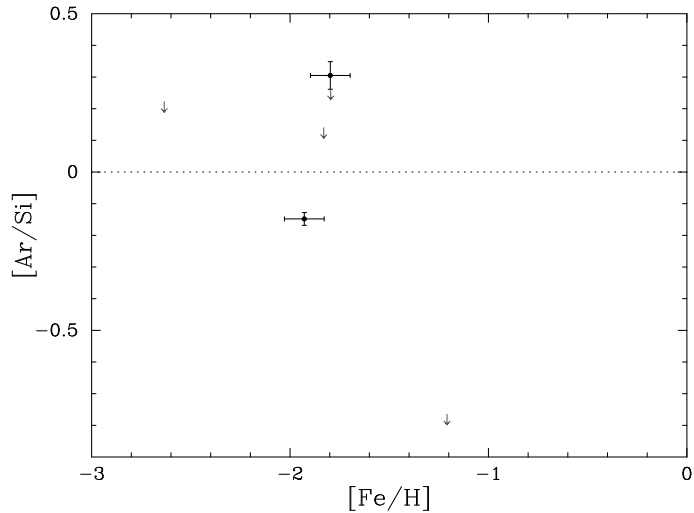}
\begin{center}
\caption{[Ar/Si] abundance ratios vs.\ [Fe/H] metallicity for our
complete sample of damped \lya systems.  
The Ar/Si ratio is an indicator of photoionization in gas-phase 
measurements with significant sub-solar values indicating photoionized
gas.  For the six damped systems with Ar observations, five are
consistent with very little photoionization, while the system toward
Q1759+75 is likely to be partially ionized.
The dashed line at [Ar/Si] = 0 indicates the solar photospheric Ar/Si ratio.
}
\label{fig:ArSi}
\end{center}
\end{figure}

Following the conclusions of
\cite{sofia98} for the ISM, \cite{molaro01} noted that an observed
solar or super-solar Ar/Si ratio significantly
limits the level of photoionization in a given damped \lya system.
This is due to the fact that 
Ar is far more sensitive to photoionizing radiation than Si.
Applying this axiom to the systems presented in Figure~\ref{fig:ArFe},
we infer that
the majority have low levels of photoionization (Figure~\ref{fig:ArSi}).
For the system toward Q1759+75, however, we note an an upper limit on Ar/Si 
which is significantly sub-solar.  We interpret this value
as an indication of substantial
photoionization and discuss the implications in a future
paper. 

\subsection{Light Elements}

\begin{figure}[ht]
\includegraphics[height=3.8in, width=2.8in,angle=-90]{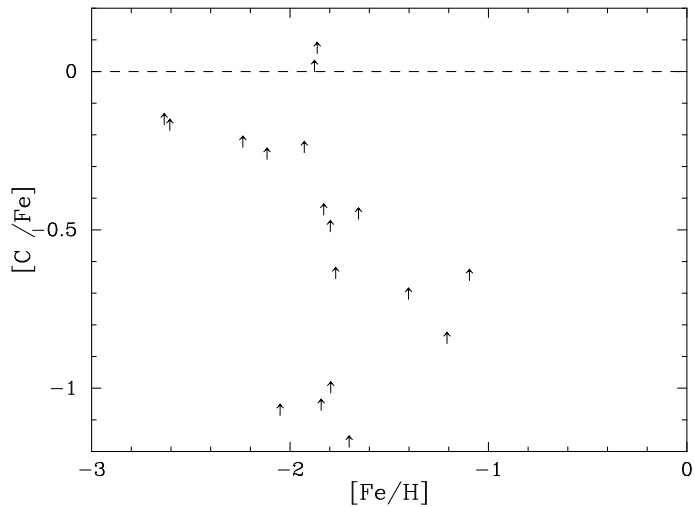}
\begin{center}
\caption{[C/Fe] abundance ratios vs.\ [Fe/H] metallicity for our
complete sample of damped \lya systems.  
We only report lower limits on C/Fe, the majority of which are too low
to provide meaningful constraints on nucleosynthesis or dust depletion.
The two systems with [C/Fe]~$>0$, however, are suggestive of significant
dust depletion and/or photoionization.
The dashed line at [C/Fe] = 0 indicates the solar photospheric C/Fe ratio.
}
\label{fig:CFe}
\end{center}
\end{figure}

\subsubsection{Carbon}

Because [C/Fe] is approximately solar in the majority of Galactic
stars (Carretta, Gratton, \& Sneden 2000; but see Wallerstein et al.\ 1997)
and because carbon is mildly refractory,
it can place unique constraints on interpretations of 
the damped \lya abundance patterns. 
Specifically, carbon can act as a tracer of the Fe-peak which is
unaffected by dust depletion.
It is very unfortunate, then, that the two UV C~II transitions are 
typically too strong to yield a meaningful limit on C/Fe.
%
Figure~\ref{fig:CFe}
plots all of the C/Fe lower limits from our observations.
We note two systems with [C/Fe]~$> 0$: the system toward
BRI0952--01 and the system at $z_{abs}=2.154$
toward Q2359$-$02.    In the case of the system toward BRI0952--01,
the C~II 1334 profile spans over 300 km/s and more resembles the C~IV
profiles than most other damped \lya systems \citep{wol00a}.  
Because C$^+$ is present
in partially ionized regions, we are concerned that this C~II
profile derives from gas in multiple phases.  Unfortunately we do
not have coverage for enough other low-ion profiles (e.g. Si~II, Al~II)
to more accurately assess this presumption.
For the system toward Q2359--02, 
the S/N of the C~II 1334 profile is poor, but
the profile is heavily saturated and [C/Fe]~$> +0.2$~dex
is possible.  In this case, the C/Fe ratio may indicate dust depletion.



\begin{figure}[ht]
\includegraphics[height=3.8in, width=2.8in,angle=-90]{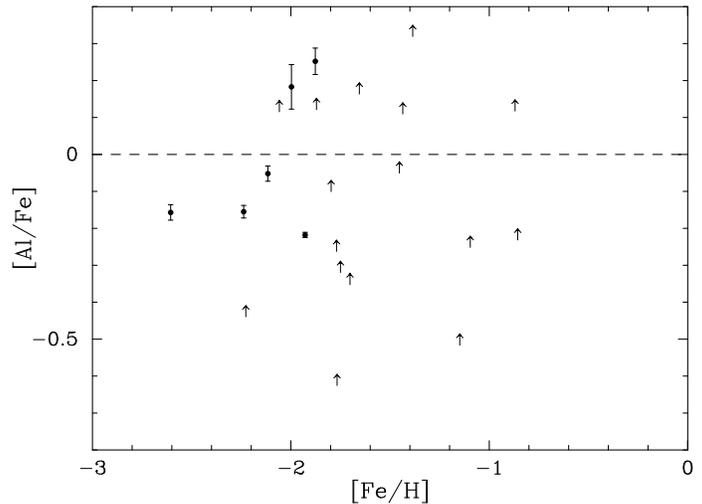}
\begin{center}
\caption{[Al/Fe] abundance ratios vs.\ [Fe/H] metallicity for our
complete sample of damped \lya systems.  
We report lower limits for the majority of systems including several
super-solar values.   Although it is difficult to assess a trend with
metallicity, 
we contend that there is an increase in Al/Fe with increasing [Fe/H].
The dashed line at [Al/Fe] = 0 indicates the solar meteoritic Al/Fe ratio.
}
\label{fig:AlFe}
\end{center}
\end{figure}

\begin{table*}[ht]
\begin{center}
\caption{
{\sc ELEMENT OVERVIEW\label{tab:Sumelm}}}
\begin{tabular}{lccccccl}
\tableline
\tableline
Element & Nucleo. Type & \multispan{2}{\hfill Galactic Stellar [X/Fe]\hfill} & 
\multispan{3}{\hfill Galactic Dust [X/Fe] \hfill} & Observing in DLA \\
& & [Fe/H]~$= -2$ & [Fe/H]~$=-1$ & Nature & 
(Halo) & (Disk) \\
\tableline
C & Light element    & 0.0  & 0.0  & mildly refractory & +0.5 & +1.7 
       & Saturated transtisions\\
O & $\alpha$ element & +0.4 & +0.4 & mildly refractory & +0.5 & +1.7 
       & Saturated transitions\\
Al& Odd-Z element    & --1.0& 0.0  & refractory  &  0.0 &  0.0 
       & Very strong transition\\
Si& $\alpha$ element & +0.3 & +0.3 & mildly refractory & +0.3 & +1.0 \\
P & Odd-Z element    & N/A  & N/A  & mildly refractory & +0.5 & +1.5 
       & Weak, \lya Forest \\
S & $\alpha$ element & +0.5?& +0.3?& non-refractory  & +0.6 & +2.0 \\
Ar& $\alpha$ element?& N/A  & N/A  & mildly refractory & +0.5 & +1.7 
       & \lya Forest\\
Ti& $\alpha$ element & +0.4 & +0.4 & refractory      & 0.0  & --0.8 
       & Weak transitions\\
Cr& Fe-peak element  & --0.3& 0.0  & refractory      & +0.1 & 0.0 \\
Co& Fe-peak element  &  +0.3& 0.0  & refractory      & ?  & 0.0   \\
Ni& Fe-peak element  &   0.0& 0.0  & refractory      & 0.0  & -- 0.2?  \\
Zn&  ?               &  +0.2& +0.1 & mildly refractory & +0.6 & +1.4 \\
\tableline
\end{tabular}
\end{center}
\end{table*}

\subsubsection{Aluminum}
\label{subsec:AlFe}

Al~II 1670 is the strongest metal-line transition 
(highest $\lambda f$-value) which we
observe in the damped \lya systems.  For this reason, it can provide
a metallicity diagnostic for even the most metal-poor systems.
In the majority of systems, however, the profile is heavily saturated
and we place only a lower limit on the Al/Fe ratio.  Figure~\ref{fig:AlFe}
presents all of the [Al/Fe] measurements and lower
limits against metallicity.  At [Fe/H]~$< -2$~dex, several of the
damped \lya systems show deficient Al/Fe ratios. In contrast, 
there are a number of lower limits with
[Al/Fe]~$\gtrsim 0$ at [Fe/H]~$>-1.5$ including 
several where the Al~II profile is heavily saturated.   In these cases the 
[Al/Fe] values are probably at least +0.3~dex.
Although the refractory nature of Al is not well established \citep{howk99c},
we note that the Al/Fe
enhancements at [Fe/H]~$\gtrsim -1.5$ and the underabundance
of Al/Fe at [Fe/H]~$< -2$ follow the general trends observed for
metal-poor stars \citep{mcw97}.

\subsection{Summary} 
\label{sec:absum}

In Table~\ref{tab:Sumelm}, we present a summary of all of the elements
analyzed in the damped \lya systems from our database
to help the reader negotiate 
the following sections on dust depletion and nucleosynthesis.  Column~1
lists the element, column~2
describes the dominant nucleosynthetic characteristic, columns~3 and 4
give the approximate [X/Fe] values in Galactic halo stars with 
[Fe/H]~$\approx -2$ and --1~dex respectively, columns 5--7 detail the 
refractory nature of each element and present rough [X/Fe] values 
in the warm halo and cool disk gas of the Galactic ISM, and
column~8 comments on any difficulties associated with measuring the 
element in damped \lya systems.

Table~\ref{tab:absum} provides a summary of the abundance measurements
for the entire sample.  Column 1 lists element X, column 2 details
the number of systems ($n$) with measurements of X, column 3 gives the mean
[X/Fe] ratio $\bar x$, column 4 is the scatter in the mean 
$\sigma(\bar x)$, column 5 presents the median [X/Fe] value $\tilde x$, 
column 6 lists Pearson's $r$-value, and
column 7 gives the probability of the null hypothesis (i.e.\ no linear
correlation) from the Pearson test. 

\begin{table}[ht]
\begin{center}
\caption{ {\sc 
ABUNDANCE SUMMARY\label{tab:absum}}}
\begin{tabular}{lrrrrrr}
\tableline
\tableline
X& $n$ & $\bar x$\tablenotemark{a} & $\sigma(\bar x)$
& $\tilde x$\tablenotemark{b} & $r_{Pears}$ & $P_{Pears}$ \\ 
\tableline
C     & 0 & ... & ...     \\
O  &   1 & $ 0.211 $ & ...    \\
Al &   6 & $-0.025$ & 0.196 & $-0.103$ & $ 0.513$ & 0.298   \\
Si &  26 & $ 0.379$ & 0.166 & $ 0.320$ & $ 0.305$ & 0.130   \\
P  &   1 & $ 0.198 $ & ...    \\
S  &   6 & $ 0.354$ & 0.138 & $ 0.330$ & $ 0.596$ & 0.212   \\
Ar &   3 & $ 0.425$ & 0.333 & $ 0.421$ & $ 0.854$ & 0.348   \\
Ti &   5 & $ 0.105$ & 0.254 & $ 0.032$ & $ 0.187$ & 0.764   \\
Cr &  16 & $ 0.166$ & 0.147 & $ 0.154$ & $-0.177$ & 0.512   \\
Co &   1 & $ 0.254 $ & ...    \\
Ni &  25 & $ 0.057$ & 0.135 & $ 0.094$ & $ 0.462$ & 0.020   \\
Zn &  13 & $ 0.450$ & 0.217 & $ 0.446$ & $ 0.014$ & 0.963   \\
\tableline
\end{tabular}
\end{center}
\tablenotetext{a}{Mean}
\tablenotetext{b}{Median}
\end{table}
 
\begin{figure*}
\includegraphics[height=9.0in, width=7.2in]{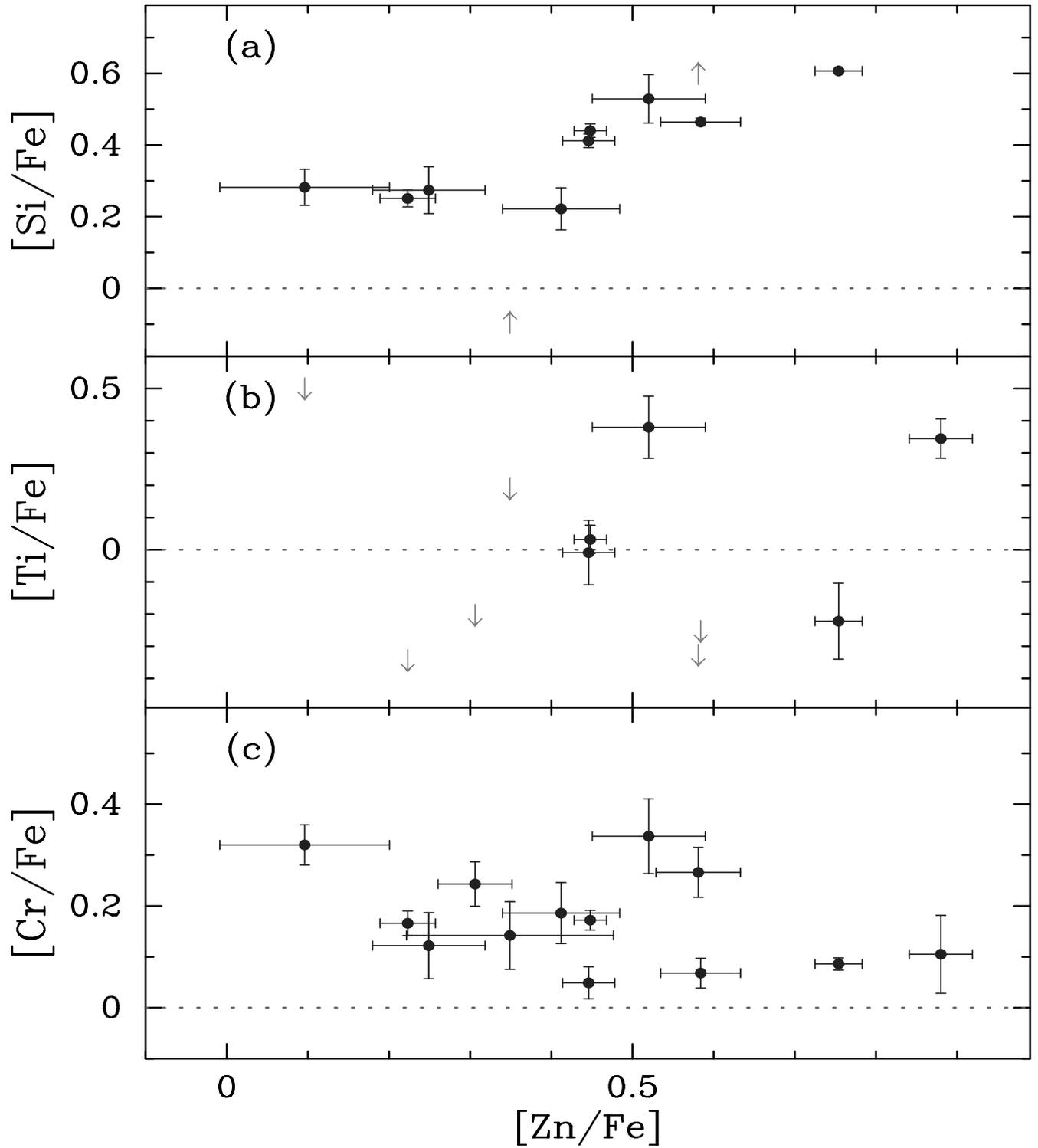}
\begin{center}
\caption{[X/Fe], [Zn/Fe] abundance ratio pairs for our sample of damped
\lya systems for (a) Si, (b) Ti, and (c) Cr.
If [Zn/Fe] enhancements indicate significant dust depletion,
one might expect trends for several of these X/Fe ratios.
}
\label{fig:XFe}
\end{center}
\end{figure*}

\section{DUST DEPLETION}
\label{sec:dust-depl}

Every sightline through the ISM exhibits evidence for the
presence of dust grains 
and there are many indications of dust in high redshift galaxies
\citep[e.g.][]{std98,richards99}.
For these reasons, one
expects that the damped \lya systems might also contain dust and 
exhibit at least modest levels of dust depletion. 
The damped \lya systems are metal-poor and
their observed abundance patterns presumably arise from the combination of
a non-solar nucleosynthetic pattern and a dust depletion
pattern.  The challenge one faces is to disentangle the two patterns in
order to
gauge the implications of dust depletion and reveal the chemical enrichment
history of these protogalaxies.

For the damped \lya systems,
the observed Zn/Fe (or  Zn/Cr) enhancements
have been identified as the most compelling argument that the refractory
elements suffer
at least modest levels of dust depletion \citep[e.g.][]{ptt94}.
Although this argument was well-founded initially, recent studies on Zn
have challenged this presumption \citep{jhnsn01,pro00,primas00,umeda01}.
In short, our empirical and theoretical
understanding of the nucleosynthesis of
Zn may be sufficiently clouded that an enhanced Zn/Fe ratio 
on its own is not a definitive argument for the presence of dust depletion.
To best address the issue of dust depletion, 
one must consider other ratios and search for additional abundance trends
(e.g. Si/Fe vs.\ Zn/Fe; PW99).
Nevertheless, examining new element ratios and trends can help break
this degeneracy.  
This is the approach we take in the following sub-sections.
To help guide the reader,
Table~\ref{tab:Sumrto} presents a summary ($\S$~\ref{sec:sumdn})
of some of the most important
abundance ratios in discussing dust depletion.

\subsection{[X/Fe] vs.\ [Zn/Fe]}
\label{subsec:XFeZnFe}

Figure~\ref{fig:XFe} presents X/Fe ratios for (a) Si, (b) Ti,
and (c) Cr against [Zn/Fe]. 
Although nucleosynthesis processes might yield Zn/Fe
enhancements of +0.1 to +0.3~dex, we contend that larger
enhancements are indicative of dust depletion.
Therefore, if one observes X/Fe departures
from the solar abundance ratio which are to be explained by 
dust depletion then one might expect a correlation (or anti-correlation)
with [Zn/Fe].  In PW99, for example,
we noted a correlation between the Zn/Fe and Si/Fe ratios 
and presented this correlation as tentative evidence for dust
depletion in the systems with the largest ratios. 
The enlarged database has
strengthened the significance of this correlation (Figure~\ref{fig:XFe}b);  
the Pearson's probability of the null
hypothesis for no linear correlation is $P_{Pears} < 0.5\%$.  
Because one observes
a similar trend along ISM sightlines, 
the correlation is suggestive of dust depletion.  

\begin{figure}[ht]
\includegraphics[height=4.8in, width=3.8in]{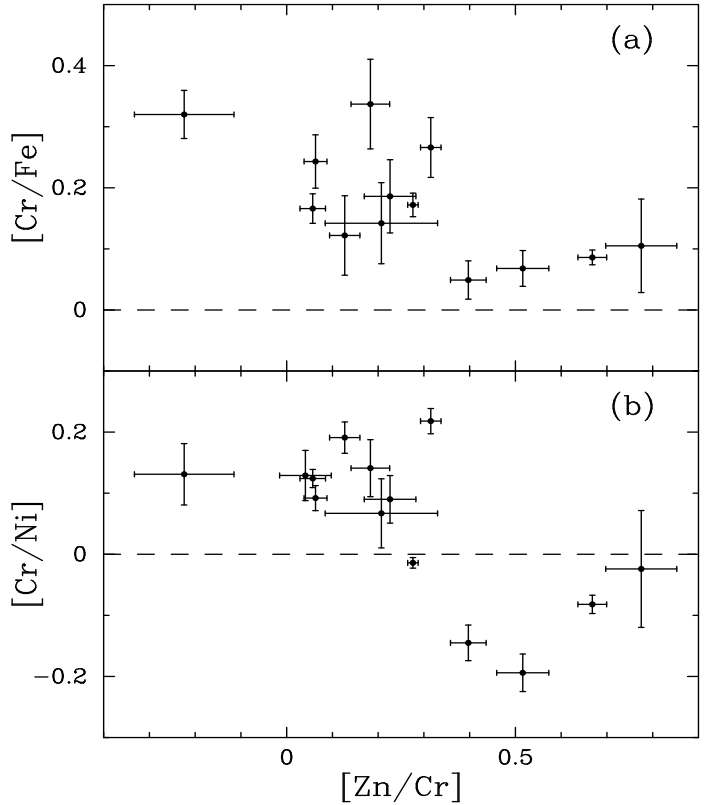}
\begin{center}
\caption{Plots of (a) [Cr/Fe] and (b) [Cr/Ni] vs.\ [Zn/Cr] for our 
complete sample of damped systems.  In both panels we note a significant
trend for decreasing [Cr/X] with increasing [Zn/Cr].  We suspect the
trends are the combination of nucleosynthesis of Cr independent of the
Fe-peak elements and dust depletion.
}
\label{fig:crznfe}
\end{center}
\end{figure}

Previous studies have noted enhancements of Cr/Fe 
in the majority of damped systems (L96; PW99).  
With the revised Fe~II oscillator strengths, the 
median enhancement is even higher: $\tilde x({\rm Cr}) = $\medcrfe~dex.  
At present, there is no stellar population which exhibits
an enhanced Cr/Fe ratio relative to solar.  In fact, the extreme metal-poor
stars tend to show deficient Cr/Fe \citep{mcw95}, 
a trend which might be explained by Type~II SN enrichment with 
a deep mass cut \citep{umeda01}.  In contrast, the
majority of Galactic ISM sightlines show super-solar Cr/Fe ratios of 
$\approx +0.05 - 0.2$~dex \citep{sav96}.  
Therefore, one would tend to interpret the damped \lya Cr/Fe enhancements
in terms of dust depletion. 
With this hypothesis, one might expect the Cr/Fe and Zn/Fe
enhancements to correlate.  In the ISM, however,
there is actually an anti-correlation where the less
depleted warm halo sightlines exhibit the largest Cr/Fe ratios and the
dense, highly depleted cold disk sightlines show solar Cr/Fe ratios
\citep{sav96}.
This trend is interpreted as occurring
because Fe is more readily 
depleted into dust grain cores while Cr is more efficiently adsorbed
into the grain mantles which form in denser, more depleted
regions.
Figure~\ref{fig:XFe}c
plots [Cr/Fe] vs.\ [Zn/Fe] for the \ncrznfe\ 
systems in our sample and reveals no
apparent trend between these two ratios: 
$r_{Pears} = -0.32, P_{Pears} = 31\%$.
We note, however, that the systems with the lowest [Cr/Fe] values
have relatively large [Zn/Fe] ratios. This point is accentuated by 
Figure~\ref{fig:crznfe} which compares (a) [Cr/Fe] and 
(b) [Cr/Ni]  against [Zn/Cr]. Here, the systems exhibit an
anti-correlation which follows
the general trend of the ISM but at 
much lower levels of dust depletion.
The quantitative difference is significant; Cr/Fe remains enhanced
in the ISM until 
[Zn/Fe]~$\approx +2.0$~dex.  This difference suggests the damped \lya
trend may not be entirely due to dust depletion.
A nucleosynthetic variation in the production of Cr {\it independent} of the 
production of Zn, Ni, and Fe could 
contribute significantly to the trends in Figure~\ref{fig:crznfe} where
Cr is reported in both the abscissa and ordinate.
Not only could a nucleosynthetic process reproduce the anti-correlation,
it could naturally explain the systems with large Cr enhancements
at [Zn/Fe]~$\approx 0$.


The Ti/Fe ratio is significantly enhanced
in the majority of metal-poor stars whereas 
Ti is often more heavily depleted than Fe in the ISM. 
Because of these competing trends, a significant offset in the
Ti/Fe ratio from the solar value would pose strong evidence for
one or the other process.
As Figure~\ref{fig:TiFe}
indicates, the damped \lya systems generally show
deficient Ti/Fe and several
cases exhibit sub-solar upper limits: [Ti/Fe]~$<-0.3$~dex.  Because these
Ti/Fe ratios contradict the empirical values for every star analyzed,
they argue for significant dust depletion.  
Values of [Ti/Fe]~$\leq -0.3$ are
generally observed only in severely dust depleted regions of the ISM 
(e.g. $\zeta$ Oph; but see Howk et al.\ 1999). 
Again, if the deficient Ti/Fe ratios are indicative of dust depletion, one might
expect an anti-correlation between [Ti/Fe] and [Zn/Fe].  Examining 
Figure~\ref{fig:XFe}b, we note no obvious anti-correlation.
Although several of
the systems with large [Zn/Fe] values show low Ti/Fe ratios, the system
with the largest [Ti/Fe] value also has a large Zn/Fe enhancement.  If the
sub-solar Ti/Fe ratios are indicative of dust depletion, then the results in
Figure~\ref{fig:XFe} suggest significant
contributions to the Ti/Fe ratios from both nucleosynthesis
and dust depletion. 

\begin{figure}[ht]
\includegraphics[height=3.8in, width=2.8in,angle=-90]{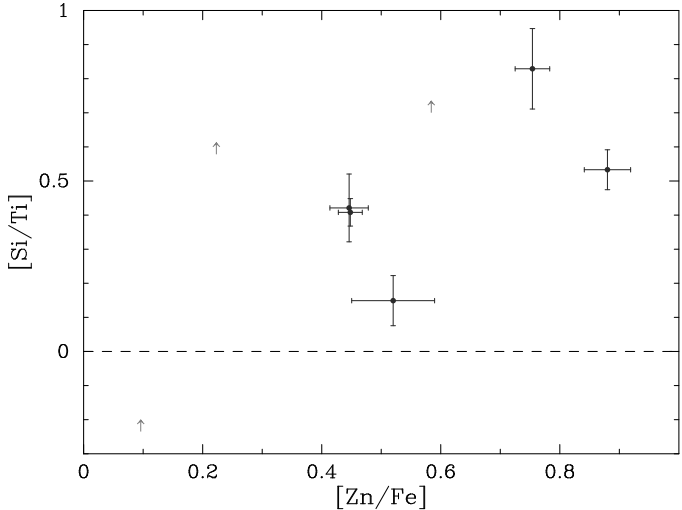}
\begin{center}
\caption{[Si/Ti] values plotted against [Zn/Fe] ratios in the few
damped systems where all four elements have been observed.
With the exception of one measurement and one lower limit, 
the systems show significantly enhanced Si/Ti and the hint of a
trend for larger [Si/Ti] at larger [Zn/Fe].  These results are in
strong conflict with the ratios observed in stars of any metallicity
and therefore provide strong evidence for dust depletion in the systems
with large [Si/Ti] and [Zn/Fe].
}
\label{fig:SiTi}
\end{center}
\end{figure}


In Figure~\ref{fig:SiTi}, we present [Si/Ti] ratios against [Zn/Fe]
for the handful of damped systems where all four elements have been observed.
The results are stunning.  All of the systems are consistent with 
enhanced Si/Ti ratios and there is the hint of increasing [Si/Ti]
with increasing [Zn/Fe].  Empirically, this presents 
{\it the most unambiguous evidence for dust depletion in our database}.  
Although the relative
abundances of the $\alpha$-elements vary in stars,
essentially no star of any metallicity shows [Si/Ti]~$> +0.2$ and none
exhibit [Si/Ti]~$> +0.3$~dex.  In contrast, the results in 
Figure~\ref{fig:SiTi} generally follow the behavior of these 
four elements due to dust depletion in the ISM. 
Although Figure~\ref{fig:SiTi} is comprised of just five measurements,
the results offer persuasive evidence for dust depletion in 
damped \lya systems with large Si/Ti ratios.
We stress that future damped \lya studies
should concentrate on providing more measurements of Ti.


\begin{figure*}
\includegraphics[height=9.0in, width=7.5in]{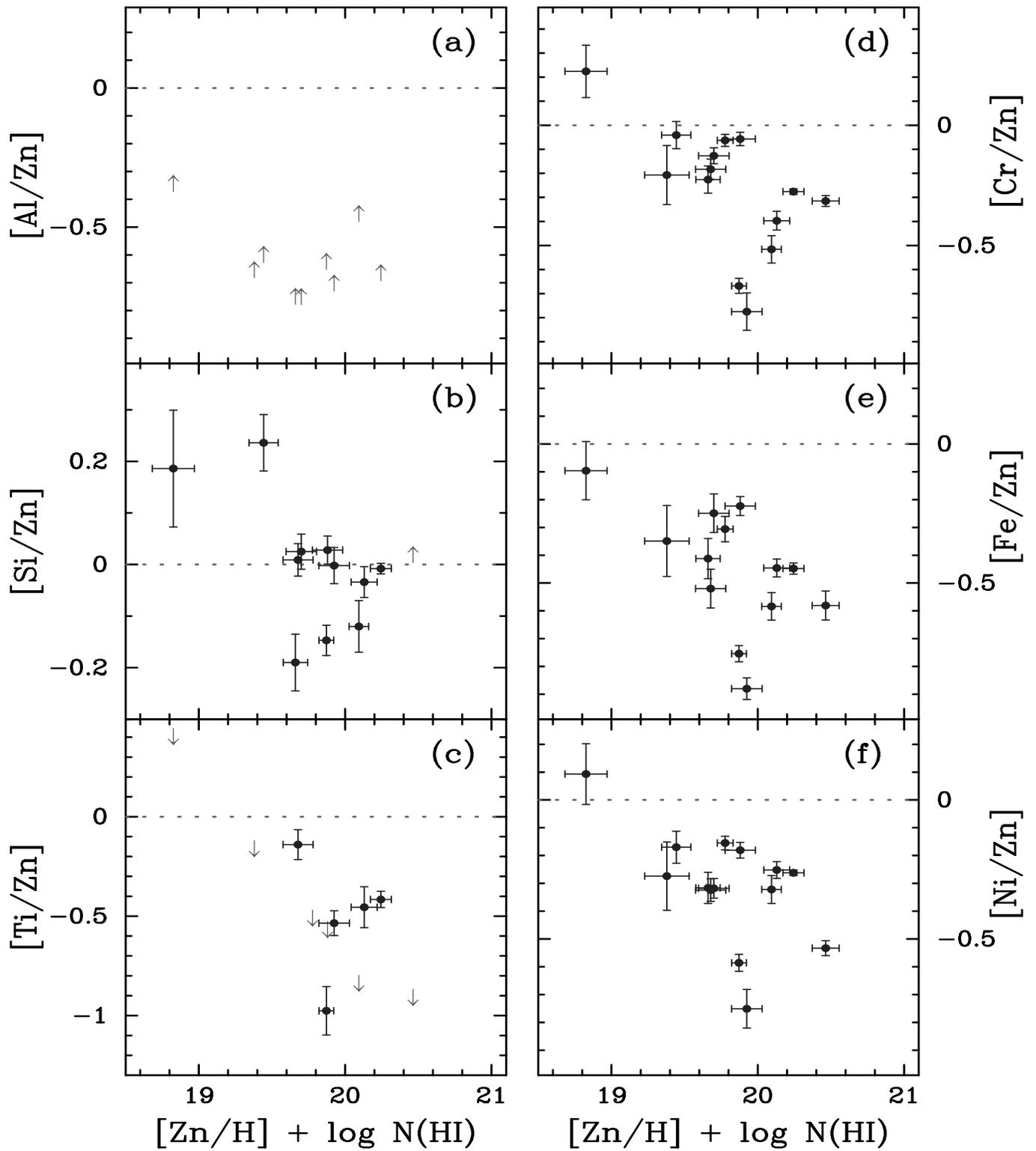}
\begin{center}
\caption{[X/Zn] vs.\ [Zn/H] + log(HI) for our complete sample of damped
\lya systems.  The majority of elements exhibit the anti-correlation
first identified by Hou et al. (2001).  These anti-correlations disappear,
however, if one removes the single system toward Q0149+33.
As noted in $\S$~\ref{sec:dust-obsc}, the absence of systems with
large [X/Zn] and [Zn/H] + log(HI) argues against dust obscuration.
}
\label{fig:XZnF}
\end{center}
\end{figure*}

\begin{figure*}
\includegraphics[height=9.0in, width=7.5in]{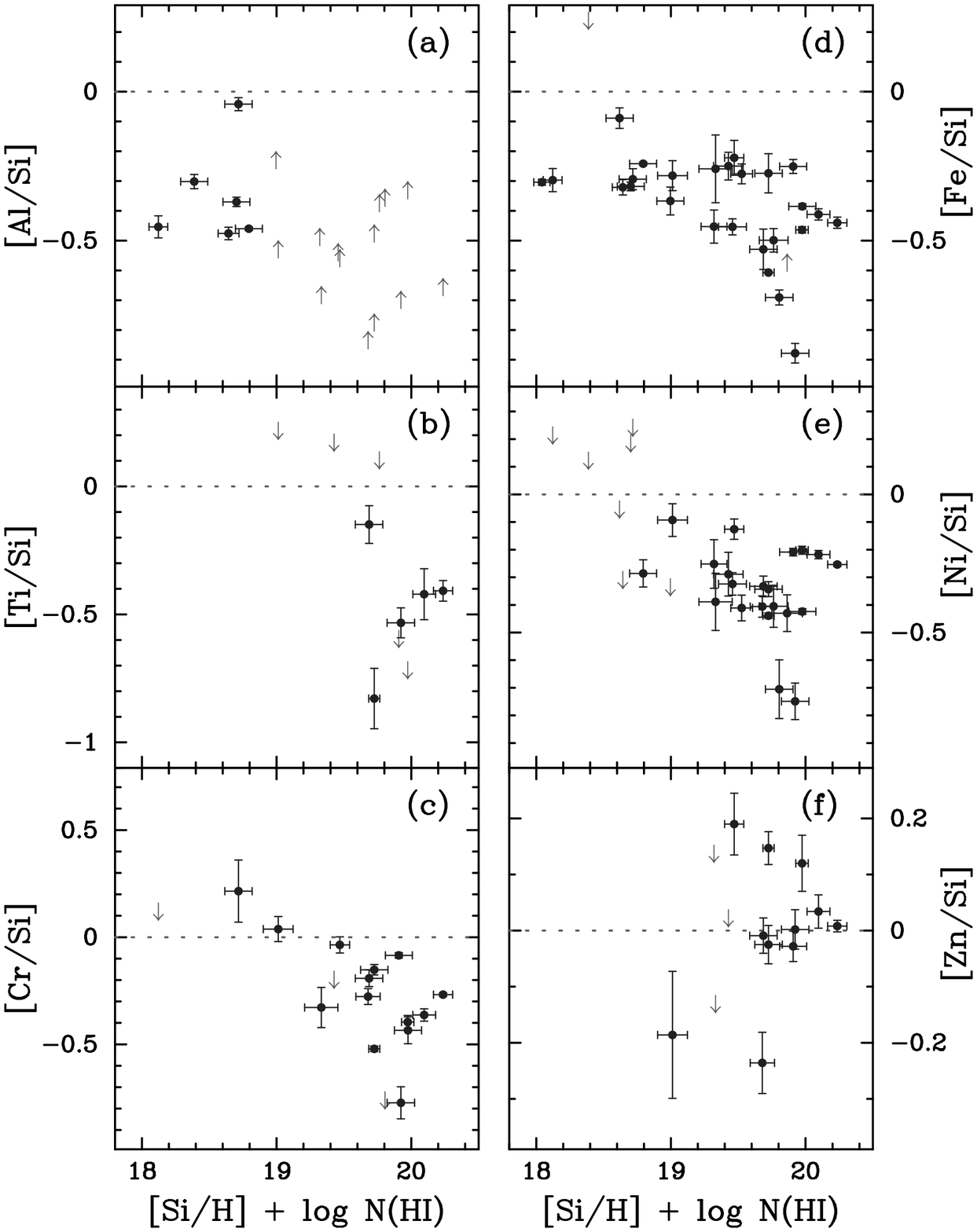}
\begin{center}
\caption{This figure is analogous to Figure~\ref{fig:XZnF} except we
have substituted the roles of Si and Zn.  In this case, we find 
statistically significant anti-correlations for almost all of the elements
which argues for dust depletion in systems with large
[Si/H] + log(HI).}
\label{fig:XSiF}
\end{center}
\end{figure*}

\subsection{[X/Y] vs.\ [Y/H] + $\log \N{HI}$}
\label{sec:hou}

In a recent paper, \cite{hou01} presented an argument for dust depletion
in the damped \lya systems
based on the anti-correlation between 
X/Zn ratios and $\F{Zn} \equiv$~[Zn/H]~$+ \log \N{HI}$ 
values which they culled from the literature.
We present these quantities for several elements in Figure~\ref{fig:XZnF}.
Because \cite{hou01} included many measurements from PW99, it is not surprising
that we find similar anti-correlations.  We worry, however,
that the observed trends might not result from dust depletion. 
As \cite{hou01} noted, an anti-correlation is observed in the ISM when 
X/Zn is considered against $\N{HI}$, presumably because 
higher $\N{HI}$ sightlines
probe denser regions \cite{wakker00}.  
In contrast, the damped \lya systems do not exhibit
trends between X/Zn and $\N{HI}$, but X/Zn and $\F{Zn}$ which is
simply $\N{Zn} + 12.0 - 4.67$.
It is not obvious that the X/Zn ratios should anti-correlate with $\N{Zn}$.
For example, sightlines with large $\N{HI}$ values but small [Zn/H]
might have very different physical properties than sightlines with high [Zn/H]
and low $\N{HI}$.  
We also have concerns that the observed 
X/Fe vs.\ $\F{Zn}$ trends could be due to variations in the
production of Zn relative to the Fe-peak elements, in particular because
the nucleosynthetic history of Zn is expected to be independent of
the Fe-peak elements \citep[e.g.][]{hff96}.  
Because the figure simply plots X/Zn vs.\ $\N{Zn}$, an anti-correlation
could arise from a significant dispersion in the production of Zn
either from galaxy to galaxy and/or within a given galaxy.
Similarly, variations in $\N{Zn}$ from uncorrected line blending
($\S$~\ref{sec:Zn}) could enhance the anti-correlations.

\begin{table}[ht]
\begin{center}
\caption{ {\sc [X/Y] vs.\ [Y/H] + log(HI)\label{tab:XYYH}}}
\begin{tabular}{lccccc}
\tableline
\tableline
Y &X & $n$ & $r_{pears}$ & $P_{pears}$ \\
\tableline
Zn & Si &  11 & $-0.600$ & 0.051\\  
& Ti &   5 & $-0.099$ & 0.874\\  
& Cr &  14 & $-0.604$ & 0.022\\  
& Fe &  13 & $-0.550$ & 0.052\\  
& Ni &  14 & $-0.582$ & 0.029\\  
\\
Zn$^a$ & Si &  10 & $-0.406$ & 0.245\\  
& Ti &   5 & $-0.099$ & 0.874\\  
& Cr &  13 & $-0.394$ & 0.183\\  
& Fe &  12 & $-0.325$ & 0.303\\  
& Ni &  13 & $-0.326$ & 0.278\\  
\\
Si & Al &   6 & $ 0.220$ & 0.675\\  
& Ti &   5 & $ 0.100$ & 0.873\\  
& Cr &  14 & $-0.666$ & 0.009\\  
& Fe &  26 & $-0.498$ & 0.010\\  
& Ni &  22 & $-0.742$ & 0.000\\  
& Zn &  11 & $ 0.322$ & 0.335\\  
\\
Si$^a$ & Al &   6 & $ 0.220$ & 0.675\\  
& Ti &   5 & $ 0.100$ & 0.873\\  
& Cr &  13 & $-0.611$ & 0.026\\  
& Fe &  25 & $-0.491$ & 0.013\\  
& Ni &  21 & $-0.736$ & 0.000\\  
& Zn &  10 & $-0.079$ & 0.828\\  
\tableline
\end{tabular}
\end{center}
\tablenotetext{a}{Statistics do not include the system toward Q0149+33}
\end{table}

Both the apparent and quantitative evidence
for the anti-correlations in Figure~\ref{fig:XZnF}
are driven by the single system toward 
Q0149+33. This system exhibits the lowest [Zn/H]~$+ \log \N{HI}$ value in
our sample and marks the detection limit of $\N{Zn^+}$ from a typical
high resolution observation.  Although we consider the relative abundance
measurements for the Q0149+33 system to be reasonably accurate, we caution
the reader against drawing significant conclusions from this single system.
To emphasize the point, we present the probability for the null hypothesis
of no linear correlation in Table~\ref{tab:XYYH} both including and
remaining the damped system toward Q0149+33.  
Although most of the elements exhibit significant anti-correlations
when Q0149+33 is included ($P_{Pears} < 5\%$), none have 
$P_{Pears} < 15\%$ when the system is removed.  Obviously,
a further examination of damped systems with very low $\N{Zn}$ values is
required to establish or contradict these
anti-correlations.  This task is difficult, however, because 
one quickly approaches the detection limit of Zn.

In lieu of new Zn measurements, we can address the anti-correlations
by considering the same analysis but replacing Zn with Si. 
Although Si is mildly refractory, the damped \lya Si/Zn
ratios are nearly solar and
do not appear to correlate with [Fe/H], $\N{HI}$, or redshift.
Therefore, we feel confident in substituting Si for Zn in this
analysis.
Figure~\ref{fig:XSiF} plots [X/Si] against $\F{Si}$ for the same
elements presented in Figure~\ref{fig:XZnF}.  For Cr, Ni, and Fe
there are statistically significant anti-correlations, none of which
hinge on a single system (Table~\ref{tab:XYYH}).  
In contrast to the analysis of Zn, the ratios do not decrease monotonically
with increasing $\F{Si}$ but instead tend to exhibit a range of [X/Si]
values at higher $\F{Si}$.  This suggests that $\F{Si}$, i.e. $\N{Si}$,
is not the dominant indicator of dust depletion.  
In the next section we argue that [Si/H] metallicity is the dominant
predictor of dust depletion in the damped \lya systems. 
Nevertheless, we
contend the general decrease in [X/Si] is the result of dust depletion,
as envisioned by Hou et al.

\begin{figure*}
\includegraphics[height=9.0in, width=7.2in]{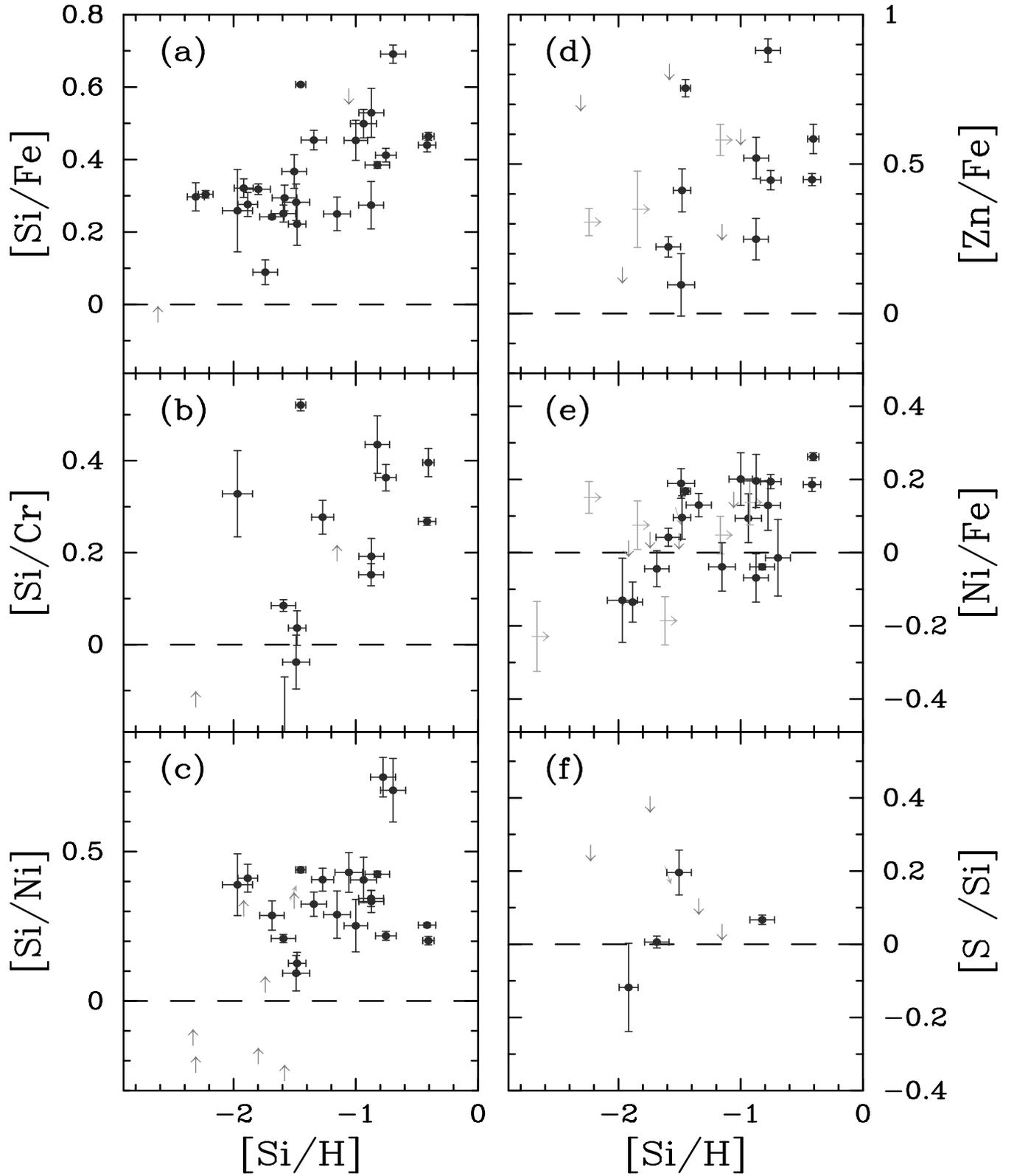}
\begin{center}
\caption{Relative abundance measurements for several select ratios
plotted against [Si/H] metallicity. The lighter shaded data points
and diagonal errors indicate those systems with lower limits to $\N{Si^+}$.
We contend that the increase in [Si/Fe] with increasing [Si/H] is indicative of 
dust depletion.  This interpretation is supported by similar trends for
[Si/Cr] and [Zn/Fe].  We also note a plateau of [Si/Fe] values at [Si/H]~$< -1.3$
which we argue is due to significant $\alpha$-enrichment in the metal-poor
damped \lya systems.
}
\label{fig:XYSi}
\end{center}
\end{figure*}

\subsection{[X/Y] vs.\ [Si/H]}
\label{sec:SiH}

In $\S$~\ref{sec:elem} we presented abundances for each element
relative to Fe against the [Fe/H] metallicity, in large part to 
facilitate comparisons with the Galactic ISM and stellar abundances.
We have concerns, however, that the effects of dust
depletion may hide significant trends with metallicity if we compare
against [Fe/H] instead of a much less refractory metallicity indicator.
To illustrate this point, we plot several abundance ratios against
[Si/H] in Figure~\ref{fig:XYSi}.  For systems with lower limits
on $\N{Si^+}$, we plot the data points as right or diagonal arrows.

The results for several of the abundance ratios are striking.  In particular,
[Si/Fe] exhibits a strong dependence on [Si/H] (Figure~\ref{fig:XYSi}a).  
The values are nearly 
constant for [Si/H]~$= -2.5$~to~$-1$~dex 
but increase by several tenths of a dex at higher metallicity.  
This trend is highly suggestive of dust
depletion in the highest metallicity systems
and argues that the plateau results from a nucleosynthetic 'floor'
(see $\S$~\ref{subsec:nuclalpha}).
Furthermore, one draws a similar conclusion from the 
Si/Cr and Zn/Fe data points
albeit at lower statistical significance.
Altogether, the trends indicate depletion levels of a few 0.1~dex in
the highest metallicity damped \lya systems.

In contrast to the Si/Fe ratios, 
the Si/Ni ratios are reasonably constant over the entire range of 
metallicity.  Furthermore, the systems with the highest two
[Si/H] values actually
exhibit two of the {\it lowest} [Si/Ni] values.  
The difference in the behavior
of Ni and Fe is difficult to understand, particularly because 
dust depletion should imply even larger Si/Ni ratios.
Although photoionization would tend to predict
enhanced Ni/Fe ratios together with higher
Si/Fe abundance, the low Ni/Cr ratios contradict this 
explanation \citep{howk99}.
Maybe the Ni abundances imply a special nucleosynthetic process which
yields super-solar Ni/Fe at high metallicity or possibly a 
dust depletion pattern different from the Galactic ISM.
The disparity between Si/Fe and Si/Ni implies that Ni/Fe increases with
[Si/H] which is partly indicated by Figure~\ref{fig:XYSi}c.
Although this trend is dominated by small number statistics,
it will be very important to pursue this with future Ni measurements.

Finally, consider the S/Si ratios presented in Figure~\ref{fig:XYSi}f.
Nucleosynthesis generally predicts a
solar relative abundance or perhaps modestly sub-solar S/Si ratios
\citep{ww95}.  On the other hand, Si is readily locked into silicate
dust grains and even the lightly depleted regions of the 
ISM exhibit [S/Si]~$> +0.2$ \citep[e.g.][]{sav96}.  
With the ISM as our guide, it appears very unlikely that
Fe would be locked into grain cores without at least
a modest depletion of Si.  Examining Figure~\ref{fig:XYSi}f, we note that
none of the systems exhibit significantly super-solar S/Si ratios.  
By chance, all of the systems with S measurements have [Si/Fe]~$\approx +0.3$
which matches the nucleosynthetic threshold implied by Figure~\ref{fig:XYSi}a.
Together the Si/Fe and S/Si ratios argue that very little Si is 
locked up in dust grains in these low [Si/H] damped systems and that the Si/Fe
and S/Fe enhancements are simply the product of nucleosynthesis.
Future S measurements of 
the significantly depleted damped systems (i.e. [Si/Fe]~$> +0.5$)
will allow for an investigation into the formation of silicates in the
damped \lya systems.

%
%
%
%


\section{NUCLEOSYNTHESIS}
\label{sec:nucleo}

For the same reasons it is difficult to identify 
abundance ratios which establish or rule out the presence of dust
depletion in the damped \lya systems, it is also challenging to reveal
the underlying nucleosynthetic patterns.  Not surprisingly, previous authors
have offered competing viewpoints.  L96 argued that the enhanced Si/Fe
and S/Fe ratios and the sub-solar Mn/Fe ratios provide compelling evidence
for Type~II SN enrichment.  At the time, they explained the enhanced Zn/Fe
ratios in terms of nucleosynthetic enrichment which -- at 
least in part -- has been borne out by recent 
theoretical and empirical studies.
In contrast, other authors have noted that if
one corrects for dust depletion under the assumption that Zn/Fe is an
accurate assessment of depletion, then the underlying 
nucleosynthetic pattern is more suggestive of solar abundances, i.e.,
enrichment dominated by Type~Ia SN \citep{kulk97,vladilo98}.
As we discussed in PW99 and P00, the two viewpoints are difficult to reconcile,
particularly because
one observes enhanced Zn/Fe ratios in metal-poor stars.
Until there is a major breakthrough in our understanding
of dust depletion in the damped systems, the most fruitful approach is to
consider abundance ratios which break the dust/nucleosynthesis degeneracy
and/or damped systems where the abundance ratios suggest negligible dust
depletion (Pettini et al.\ 1999; PW99; 
Molaro et al.\ 2000).  
We adopt this method in the following subsections, stressing
several abundance patterns which reflect on nucleosynthesis largely
independent of dust depletion.
A brief summary of the most important points is presented in 
Table~\ref{tab:Sumrto} ($\S$~\ref{sec:sumdn}).

\clearpage

\subsection{Investigating Type~I vs.\ Type~II SN Enrichment 
with [X/Fe] Ratios}
\label{subsec:nuclalpha}


Si is the only $\alpha$-element that is routinely
measured in the damped \lya systems.  
Figure~\ref{fig:SiFe} shows that every damped \lya system exhibits 
enhanced [Si/Fe] and all but one has [Si/Fe]~$\geq +0.2$~dex.
Although it is difficult to separate the competing effects of dust 
depletion and nucleosynthesis on the Si/Fe enhancements, consider the
following arguments.
In Figure~\ref{fig:XFe}a, we noted
that the Si/Fe enhancements tend to [Si/Fe]~$\approx +0.25$~dex 
as [Zn/Fe]~$\to 0$, i.e., in dust-free systems.
This trend suggests the damped \lya systems have $\alpha$-enhancements
of +0.2 to +0.3~dex although the argument is tentative because 
only a few systems have [Zn/Fe]~$\approx 0$.
Our interpretation is substantiated, however, by the results presented in 
Figures~\ref{fig:XSiF}a and \ref{fig:XYSi}a which plot
[Fe/Si] and [Si/Fe] values against $\N{Si}$ and [Si/H] respectively.
Here, the Si/Fe ratios have a nearly constant 
value of [Si/Fe]~$\approx +0.3$~dex at [Si/H]~$<-1.5$
and does not show larger values until [Si/H]~$> -1$ where dust depletion
is probably making a significant contribution.  Together these trends
imply the metal-poor damped \lya systems are primarily
enriched by Type~II SN.

A Type~II SN enrichment interpretation is further 
supported by the S/Si ratios. Because sulfur is non-refractory, 
one can use the difference in the refractory nature of S and Si
to gauge the level of Si depletion, i.e., by searching
for systematically large S/Si ratios. 
The absence of such a signature in Figure~\ref{fig:XYSi}f argues
that systems with low [Si/H] have no silicate dust grains and presumably
are essentially dust-free.  This further supports our contention that
the Si/Fe enhancements in these systems result from nucleosynthesis
and not dust depletion.

L96 and \cite{pro97a} have stressed that Mn/Fe and Ti/Fe ratios can provide 
insight into the debate on Type~II and Type~Ia 
nucleosynthesis independent of dust depletion, but recent 
developments appear to limit the impact of these elements.
First,
recent abundance studies of the Sagittarius dwarf galaxy
reveal sub-solar Mn/Fe abundances in stars with nearly
solar $\alpha$/Fe ratios \citep{smecker01} contrary
to the majority of Galactic stars.
If these results hold and are not an anomaly when compared with
abundances of other dwarf galaxies, the impact of Mn in
terms of nucleosynthesis will be reduced.
Second our additional Ti measurements and the revised Ti~II 1910
oscillator strengths indicate the majority of damped
systems have solar or sub-solar Ti/Fe values.
In these cases, it is difficult to comment on nucleosynthesis because the
low Ti/Fe ratios are likely the result of dust depletion
(e.g. Figure~\ref{fig:SiTi}).  
%
Hopefully, the next generation of echelle
spectrographs with higher UV sensitivity
will yield a number of Ar measurements and 
more meaningful lower limits on $\N{O^0}$.  These
elements are nearly non-refractory and are important diagnostics of
$\alpha$-enrichment.  In turn, comparisons among the $\alpha$-elements
provide insight into the enrichment histories of the damped
\lya systems as well as meaningful information on dust depletion. 

\subsection{Comparisons Along the Fe-Peak}
\label{subsec:nuclFepeak}

In previous sections, we noted departures from the solar
ratio for Cr/Fe and Ni/Fe which bear repeating.  Specifically, nearly
every damped \lya system exhibits super-solar Cr/Fe and the majority
show mild Ni/Fe enhancements.  In terms of Cr, we suspect that
dust depletion is playing a role in the relative abundances 
when [Zn/Fe]~$> +0.3$
($\S$~\ref{subsec:XFeZnFe}),
but note that Cr/Fe is enhanced even in systems which
are essentially dust-free (i.e. [Zn/Fe]~$\approx 0$).  Furthermore,
many of the [Cr/Fe] values are in excess of the highest values in the
ISM \citep{sav96}.  We contend these trends may have a nucleosynthetic
origin. Furthermore,
we suspect the anti-correlation observed 
between [Cr/Fe] and [Zn/Cr] (Figure~\ref{fig:crznfe}) arises from
a nucleosynthetic process. 
Although there is evidence in metal-poor stars for
some differences among the Fe-peak elements, Cr/Fe is actually 
underabundant at low metallicities while [Ni/Fe]~$\approx 0$
at all metallicity.  If these differences between the damped \lya
patterns and the Galactic stellar abundances hold, they may
indicate important nucleosynthetic processes for Ni, Cr, and Fe
which have not yet been appreciated.  They could also
reflect differences between gas-phase
and stellar atmospheric abundances produced by differential supernova
enrichment. 

At present, Cr, Ni, and Fe are the best studied Fe-peak elements
in the damped \lya systems.
There is some promise, however, to analyze Co in a number of 
systems \citep[e.g.][]{ellison01}.  It will be particularly interesting to
find if the damped systems show [Co/Fe]~$>0$~dex like metal-poor
stars and if the Co abundances track Zn as suggested by
\cite{umeda01}.

\subsection{Odd-Z Elements}
\label{sec:oddelm}

To date, aluminum has not received much attention in the damped \lya systems
except for its impact on the likelihood of photoionization
\citep[PW96;][]{howk99,vladilo01}.  This light element, however, is special
for being one of the few elements observed in damped systems
with an odd atomic number.  Determining the Al abundance
is made difficult both by
our limited knowledge of its refractory nature \citep{howk99c} and
its possible sensitivity to photoionization \citep{vladilo01}. 
Nevertheless, as noted in $\S$~\ref{subsec:AlFe},
there may be a discernible trend of [Al/Fe] with [Fe/H] metallicity.
In particular, a number of systems with [Fe/H]~$> -2$ exhibit enhanced
Al/Fe ratios while the few measurements with sub-solar Al/Fe are at 
[Fe/H]~$< -2$.  
This trend loosely follows the behavior of Al in Galactic metal-poor
stars \citep{mcw97} and may reflect a similar nucleosynthetic history
as well as a significant metallicity dependence to the production of
Al relative to Fe.

\begin{figure}[ht]
\includegraphics[height=3.8in, width=2.8in,angle=-90]{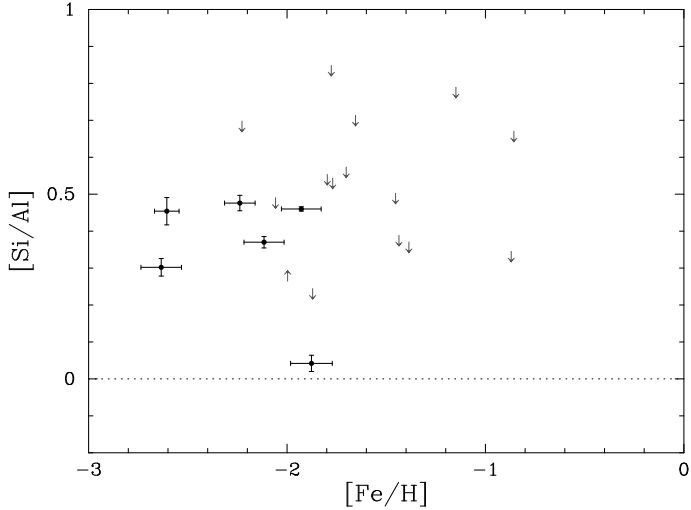}
\begin{center}
\caption{[Si/Al] values plotted against [Fe/H] metallicity.  Contrary
to predictions of nucleosynthesis and observations of Galactic metal-poor
stars, the damped \lya systems only show mildly enhanced Si/Al ratios.
Furthermore, there is no trend in the values where one might have 
expected increasing Si/Al with decreasing [Fe/H].  The results suggest
that the damped \lya systems experienced a different chemical 
enrichment history from the Galactic halo.
}
\label{fig:SiAl}
\end{center}
\end{figure}

Because Al is an odd-element, it is particularly valuable to
compare its abundance with a nearby even-element (e.g. Si)
in order to probe the so-called 'odd-even effect' \citep{arnett71}. 
This effect refers to the higher relative abundance of elements with even 
atomic numbers because of the strong coupling (and therefore greater
stability) of pairs of nucleons. 
Recent work on nucleosynthesis in zero metallicity stars indicates a
strong odd-even effect in elements with Z~$<25$, for example,
[Si/Al]~$\gtrsim 1$ \citep{heger01}.  Ideally, we could test these
predictions with measurements of damped \lya systems at [Fe/H]~$< -3$,
but it appears that no systems exist with these very low metallicities
($\S$~\ref{sec:chemev}).  
Instead, we must compare against scenarios which allow for at least
one prior generation of star formation.
Figure~\ref{fig:SiAl} plots [Si/Al] against [Fe/H] for our complete
damped \lya sample.  
Although the majority of systems have enhanced Si/Al, none
have [Si/Al]~$\approx 1$.  We emphasize
that the refractory nature of Al suggests the [Si/Al] values might
be {\it overestimates} of the intrinsic abundance.
In fact, if one assumed that the
Si/Fe enhancements are solely due to dust depletion and that
Al has the same refractory behavior as Fe, then the dust-corrected
[Si/Al] ratios would be approximately solar.

Even without a dust-correction\footnote{The effects of
photoionization might balance the
effects of dust depletion (Vladilo et al.\ 2001)},
the Si/Al ratios are lower
than the Si/Al ratios of metal-poor stars with similar metallicity
\citep{mcw95}.  
Furthermore, the damped \lya systems show no sign of increasing
Si/Al with decreasing metallicity.
In terms of Al, we contend that the
damped \lya systems have a different enrichment history than
Galactic stars with [Fe/H]~$\approx -2$.  Perhaps the differences
indicate a greater contribution of lower mass supernovae
in the damped \lya systems.  Another possibility is that the metals
observed in the damped systems are the products of metal-rich star
forming regions which have been diluted by primordial gas.

In passing, we note that one can also probe the odd-even effect
through measurements of phosphorus.
In comparison with aluminum, 
phosphorus is essentially non-refractory and should have negligible
ionization corrections.  Furthermore
the P~II 1152 transition has a smaller oscillator strength and will provide
a number of measurements at higher metallicity where the Al~II 1670
transition is always saturated.  Therefore, future abundance
studies should allow us to consider the odd-even effect over the full
range of damped \lya metallicities.

\vskip 0.2in

\subsection{'Correcting' for Dust Depletion}
\label{subsec:dustcorr}

In this section, we have focused on abundance ratios relevant to the
nucleosynthetic history of the damped \lya systems which are largely
independent of dust depletion.
At the present time, we feel there is not enough information from
the damped 
\lya observations to allow for a quantitative analysis of dust
depletion corrections.  In particular, we have serious
concerns about the assumptions of the dust depletion pattern and
overall depletion level of the damped systems.
For example, previous authors have simply assumed Zn/Fe is intrinsically
solar and constrained their dust correction analysis accordingly.
Because there may be a nucleosynthetic contribution to Zn/Fe,
this implies an overestimate to these dust corrections.
Although it is not very constructive,
instead of presenting our own analysis
we offer some criticism on treatments in the literature.

The most comprehensive approach to dust corrections in the
damped \lya systems was presented by \cite{vladilo98}.  
In his prescription, Vladilo assumed that dust
in the damped \lya systems has the same average number of atoms of each
element per grain as Galactic dust.  
If, for example, Galactic dust
has two Mn atoms depleted for every 10 Fe atoms, then 
this ratio applies for the damped \lya systems {\it irrespective 
of the total number of Mn and Fe atoms present}.  
In other words, the dust grain composition (e.g. silicates, oxides)
is independent of the relative quantities of the elements present
during grain formation.
Physically, one expects the two-body process of adhesion of
an element onto dust grains to 
be proportional to the product of the
densities of that element and the dust grains.
If there are fewer atoms available (i.e. due to a nucleosynthetic
underabundance), then the adhesion rate will be correspondingly lower.  
Aside from this assumption being non-physical, this
treatment systematically reduces the observed departure
from solar of any refractory element relative to Fe
because the corrections are inversely proportional to the
gas-phase abundances (see the Appendix).
At some level, this treatment biases the analysis
in favor of Vladilo's conclusion
that the damped \lya systems have solar relative composition.
Until someone develops a more quantitative model of dust grain formation
in the ISM, the practice of dust corrections will be
too difficult to apply in detail to the damped \lya systems.

\begin{table*}[ht]
\begin{center}
\caption{ {\sc 
RATIO COMPARISON SUMMARY\label{tab:Sumrto}}}
\begin{tabular}{llccc}
\tableline
\tableline
Ratios & Trend & Nucleosynthesis & 
Dust Depletion & Comments \\
\tableline
Si/Fe vs.\ Zn/Fe & Plateau  +& Plateau at [Zn/Fe]~$< +0.4$ 
          & Corr. at [Zn/Fe]~$>+0.4$ \\
 &Correlation&argues for $\alpha$-enrichment & argues for Fe depletion \\
Cr/Fe vs.\ Zn/Fe & Anti-correlation & Suggests nucleosynthesis & Might
	be consistent with \\
 &&&depletion\\
Ti/Fe vs.\ Zn/Fe & No trend & Large Ti/Fe value at low & Low Ti/Fe values 
		at high \\
 && Zn/Fe implies $\alpha$-enrichment & Zn/Fe implies depletion\\
Si/Ti vs.\ Zn/Fe & Correlation & & Strong evidence for the \\
&&&depletion of Ti
\\
Si/Fe vs.\ [Si/H]& Plateau + & Plateau at [Si/H]~$< -1.5$ & Corr. at
        [Si/H]~$>-1$ \\
 & Correlation & argues for $\alpha$-enrichment & argues for Fe depletion \\
Ni/Fe vs.\ [Si/H]& No trend & & & Mild enhancement difficult \\
 &&&&to account for\\
Si/Cr vs.\ [Si/H]& Correlation & & Argues for depletion of Cr \\
Si/Ni vs.\ [Si/H]& No trend & & Contradicts Si/Fe, Si/Cr? & Photoionization?\\
S/Si  vs.\ [Si/H]& No trend & & Solar value implies no depletion \\
\\
Si/Al vs.\ [Fe/H]& No trend & Low Si/Al value \\
 &&contradicts halo stars \\
\tableline
\end{tabular}
\end{center}
\end{table*}

\subsection{Summary of Dust Depletion and Nucleosynthesis}
\label{sec:sumdn}

This paper presents discussion and comparison on many relative 
abundance ratios for the damped \lya systems to develop arguments
for various aspects of dust depletion, nucleosynthesis, etc.
In Table~\ref{tab:Sumrto}, we present an executive summary of these
important abundance ratios.


\section{DUST OBSCURATION}
\label{sec:dust-obsc}

The prospect that extragalactic objects are obscured by dust
was first stressed by \cite{ostriker84} and then thoroughly investigated 
for the damped \lya systems by
\cite{fall93}.  The obvious concern regarding dust obscuration is
that systems with significant dust extinction (e.g. highly depleted,
high $\N{HI}$ systems) may drop from an optically selected quasar sample
and therefore bias surveys against large $\N{HI}$ and/or possibly high
metallicity systems \citep[e.g.][]{boisse98}.  
In this section, we consider several diagnostics for probing
the likelihood and effects of dust obscuration.

\begin{figure}[hb]
\includegraphics[height=4.5in, width=3.8in]{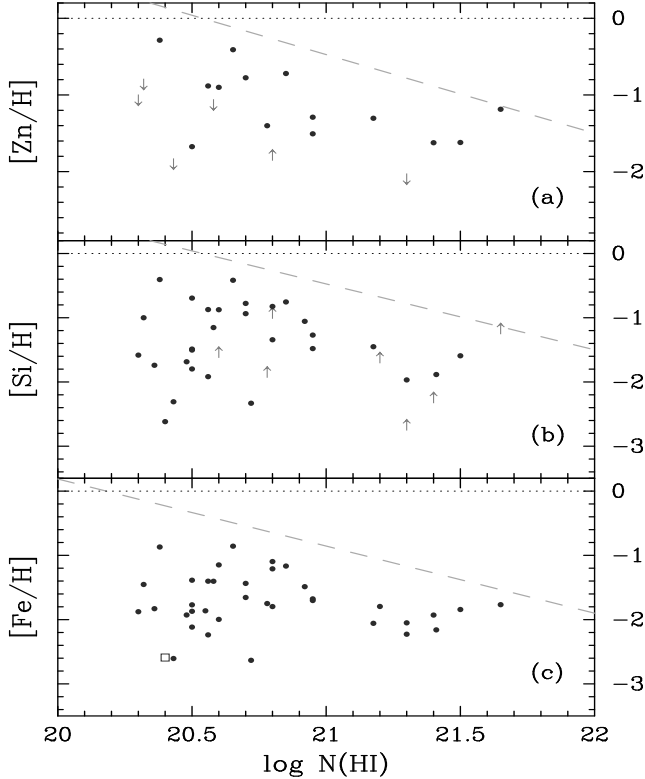}
\begin{center}
\caption{Plot of (a) [Zn/H], (b) [Si/H], and (c) [Fe/H] vs.\ $\N{HI}$.  
Boiss$\acute e$ et al.\ (1998) and others have argued that the absence
of measurements in the upper-right corner of each sub-panel is 
consistent with dust obscuration.  While our observations support the
trend identified by Boiss$\acute e$, we warn that they do not require
the interpretation of dust obscuration.
}
\label{fig:boisse}
\end{center}
\end{figure}

\subsection{[X/H] vs.\ $\N{HI}$}

Following the analysis of \cite{boisse98}, we can examine 
metallicity trends with HI column density 
to search for evidence of dust obscuration.  These authors
identified an anti-correlation between [Zn/H] and $\N{HI}$ noting that the
observed damped systems are constrained to have [Zn/H], $\N{HI}$ values
in the following range: \\
$19.8 < $~[Zn/H]~$+\log \N{HI} < 20.5$.
The lower bound is simply the observational limit for
detecting Zn, particularly in lower resolution observations 
\citep[e.g.][]{ptt94}.  
Our database includes several upper limits and measurements
below this lower threshold and the observed
[Fe/H] values suggest many more systems. 
Boiss$\rm \acute e$ et al.\ interpreted the upper bound, 
meanwhile, as the result of dust
obscuration, i.e., QAL surveys are biased
against high metallicity, high $\N{HI}$ systems. 
We shall refer to this upper threshold as the 'obscuration
threshold' for the remainder of the paper.

In Figure~\ref{fig:boisse} we plot [X/H] values for three elements drawn
from our database: (a) Zn, (b) Si, and (c) Fe.  Overplotted as a dashed
line in
each panel is the obscuration threshold adopted by \cite{boisse98} with
an offset of $-0.4$~dex for Fe to account for the observed enhancement of Zn/Fe. 
All of these elements generally follow the obscuration threshold,
although one system (Q0458--02; the rightmost data point) probably
exceeds the bound for Si.
We stress that a strict obscuration threshold -- as adopted
by several authors -- is both non-physical and apparently 
inconsistent with our observations.  Instead, one should implement 
a more realistic treatment like the one
introduced by \cite{fall93} to accurately assess the effects of
dust obscuration.

Although dust obscuration might explain the distribution of 
[X/H], $\N{HI}$ measurements presented in 
Figure~\ref{fig:boisse}, it does not {\it establish}
that damped \lya systems exhibit significant dust extinction.
The absence of high [X/H], high
$\N{HI}$ systems could very easily be explained by higher metallicity
regions having significantly consumed HI gas and therefore lower
HI column densities.  
In fact, one observes a mild anti-correlation between gas content
and stellar metallicity 
in present-day spiral galaxies \citep{zaritsky94}.
Therefore, systems with large \N{HI} column 
densities might be more likely to reflect regions where star formation has
yet to proceed in earnest.  The absence might also reflect the 
detailed morphology of the damped \lya systems.  For example, the
inner regions of protogalaxies might have HI 'holes' analogous to
the holes observed in modern spiral galaxies including our own
Galaxy \citep{vandriel87,wol98}.
Conversely, low $\N{HI}$ sightlines which show a range of metallicities,
might probe both heavily enriched regions 
and the outer, less enriched regions of protogalaxies. 
To distinguish between these scenarios, one must
consider more direct measures of dust extinction.

\begin{figure*}[ht]
\includegraphics[height=3.8in, width=2.8in,angle=-90]{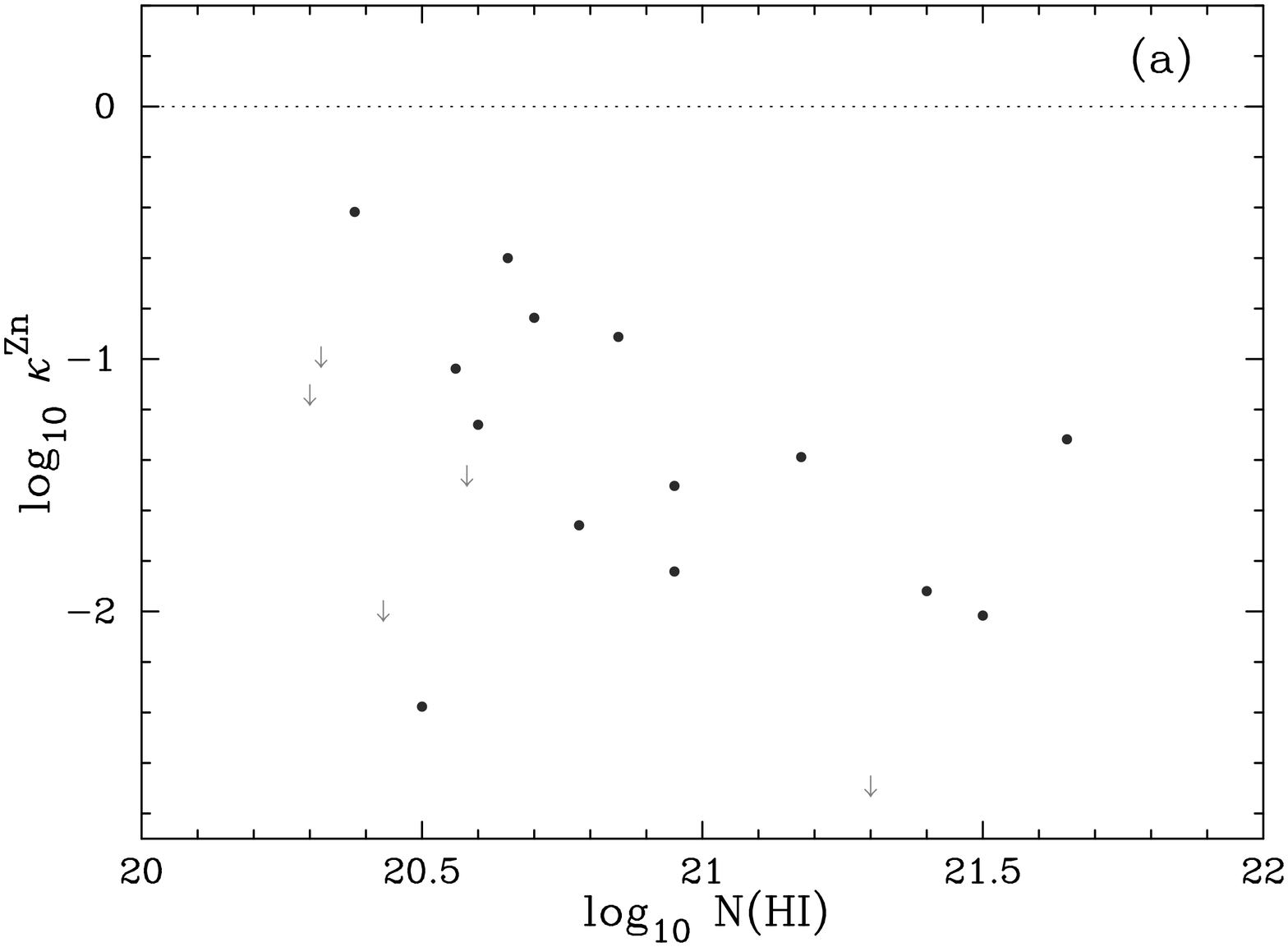}
\includegraphics[height=3.8in, width=2.8in,angle=-90]{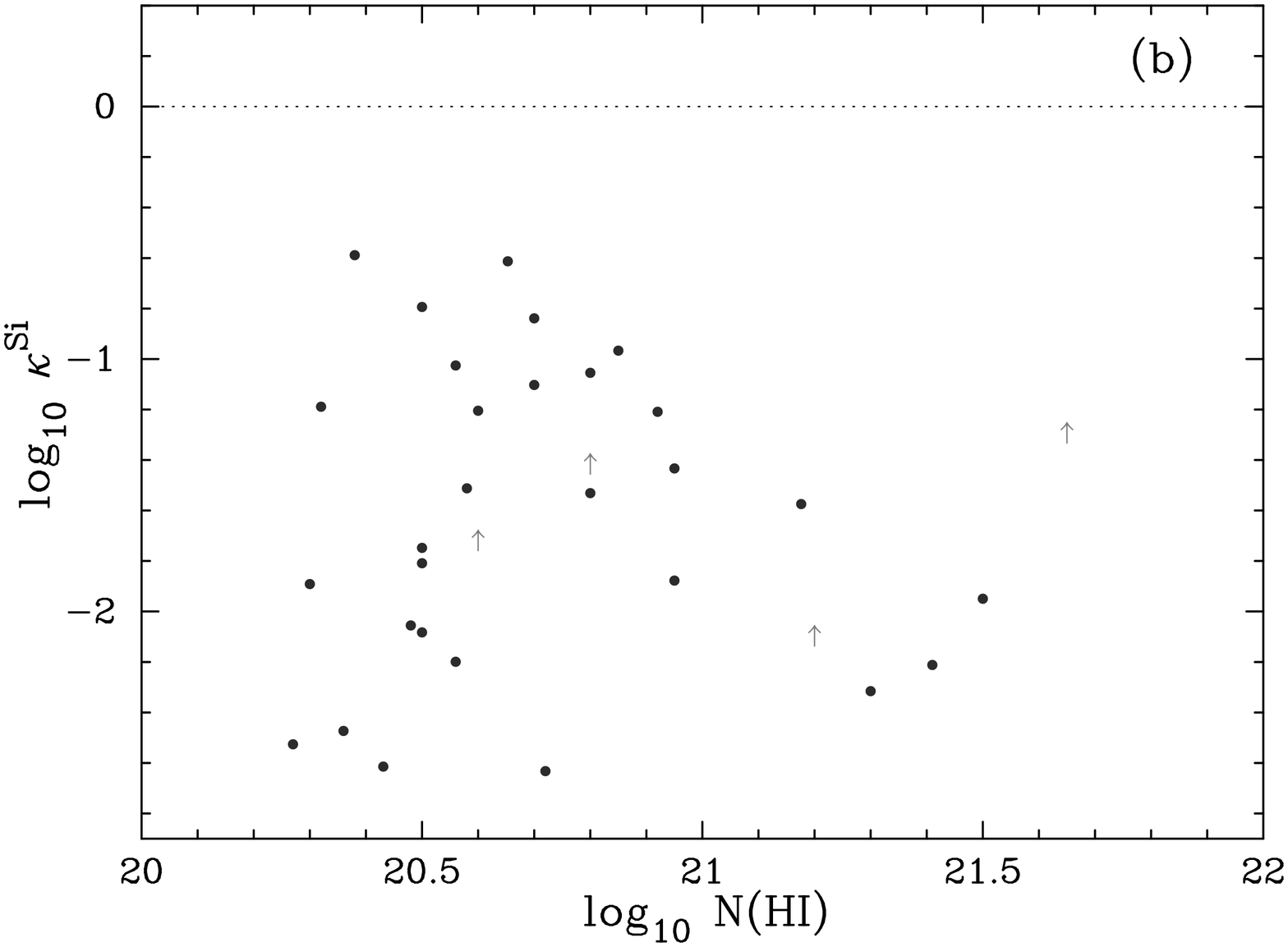}
\begin{center}
\caption{Plot of the dust-to-gas ratio $\kappa$ vs. $\N{HI}$ inferred from
the observed (a) Zn/Fe and (b) Si/Fe ratios.  There is
an absence of systems with large dust-to-gas ratios and large
$\N{HI}$, which is consistent with -- but not proof for --
dust obscuration.  In each case, the dotted line roughly corresponds
to the dust-to-gas ratio of the ISM.
}
\label{fig:dustgas}
\end{center}
\end{figure*}

\subsection{Dust-to-gas ratio vs.\ $\N{HI}$}

A more physically motivated approach 
for investigating dust obscuration
than plotting [X/H] vs.\ $\N{HI}$ 
is to consider the trend of dust-to-gas ratios $\kappa$ with HI column
density.  If dust obscuration is important,
then large $\N{HI}$ systems should have
lower dust-to-gas ratios independent of metallicity.
We can calculate dust-to-gas ratios relative to the Galactic ISM
by assuming the observed
Zn/Fe enhancements are entirely due to dust depletion, i.e.,

\begin{equation}
\kappa^{Zn} \equiv (1 - 10^{-{\rm[Zn/Fe]}}) \, 10^{ {\rm [Zn/H]}} \perd
\label{eqn:kappazn}
\end{equation}
Figure~\ref{fig:dustgas}a
presents $\kappa^{Zn}$ values for the \nZn\ systems exhibiting Zn in our
full sample.  Because [Zn/Si]~$\approx 0$ in these systems
(the median [Si/Zn] value is \medsizn), we can also consider the dust-to-gas
ratios for Si,
\begin{equation}
\kappa^{Si} \equiv (1 - 10^{-{\rm[Si/Fe]}}) \, 10^{ {\rm [Si/H]}} \cmma
\label{eqn:kappasi}
\end{equation}
which allows us to include many more systems (Figure~\ref{fig:dustgas}b).
We warn the reader, however, that a significant fraction of the observed
Si/Fe (and Zn/Fe) enhancements might be the product of nucleosynthesis,
not dust depletion.  In this case, the values presented in 
Figure~\ref{fig:dustgas} must be considered {\it upper limits} to $\kappa$.
In each sub-panel, 
the dotted line at $\log \kappa = 0$ roughly corresponds to the
Milky Way dust-to-gas ratio where the majority of refractory elements
are locked up in dust grains.

The damped \lya
dust-to-gas ratios follow the trends observed in Figure~\ref{fig:boisse};
the lower $\N{HI}$ sightlines exhibit a large range in $\kappa$ while the
higher $\N{HI}$ sightlines show only moderate to low values.
The similarity between 
Figures~\ref{fig:boisse} and \ref{fig:dustgas} stems from the fact that
the observed [Si/Fe] and [Zn/Fe] values
are only mildly dependent on metallicity and $\N{HI}$.
Like Figure~\ref{fig:boisse}, the observed distribution of
$\kappa, \N{HI}$ pairs supports the notion that a sample of
dusty, high $\N{HI}$ systems may be absent in our
sample of damped \lya systems.  

\subsection{Implications for Extinction}
\label{sec:extinct}

For dust obscuration to be important in damped \lya surveys,
the systems must contain enough dust to imply significant levels of
dust extinction.
%
Our approach is to combine the dust-to-gas ratios from the previous
sub-section with the known $\N{HI}$ values and an assumed extinction
curve to estimate 
extinction corrections for a sub-set of
our damped \lya sample. 
To calculate the extinction $A(\lambda)$ at a given wavelength
$\lambda$, we adopt an extinction curve 
$\xi(\lambda) \equiv A(\lambda)/A(V)$ and assume that 
\begin{equation}
A(V) = 3.1 \; \cdot \; \frac{\N{HI}}{5.9 \sci{21} \cm{-2}} \; \cdot \; \kappa
\end{equation}
This is equivalent to assuming that $R_V \equiv A(V)/E(B-V) = 3.1$
and $\N{HI} = 5.9 \sci{21} \cm{-2} E(B-V)$ \citep{bohlin78}.
Inherent to this prescription is the presumption that the dust-to-gas
ratio inferred from the depletion of Fe (i.e.\ $\kappa^{Si}$)
corresponds to the dust-to-gas ratio of the dust responsible for extinction.  
Because the extinction in the FUV is dominated by small dust grains
like silicates and iron oxides, the Fe depletion level should be a
reasonably good tracer of the dust responsible for dust extinction in
the damped \lya systems.
On the other hand, we again warn the reader that the probable
nucleosynthetic contribution to the observed Si/Fe ratios 
implies the $\kappa^{Si}$ values are strictly upper limits.
We consider two extinction curves which hopefully bracket that 
for the damped \lya systems: (1) MW -- the Milky
Way extinction law with $R_V = 3.1$ \citep{cardelli89}; 
and (2) SMC -- the SMC extinction law \citep{prevot84} with
an extrapolation of the form $\xi(\lambda) \propto \lambda^{-1}$
at small $\lambda$. 
Finally, we limit the analysis to quasars 
discovered in surveys with reasonably well-defined filters $X$ and
limiting magnitudes $m_{lim}$ \citep[e.g.][]{wol86,storr96}. 

\begin{figure}[ht]
\includegraphics[height=5.0in, width=3.8in]{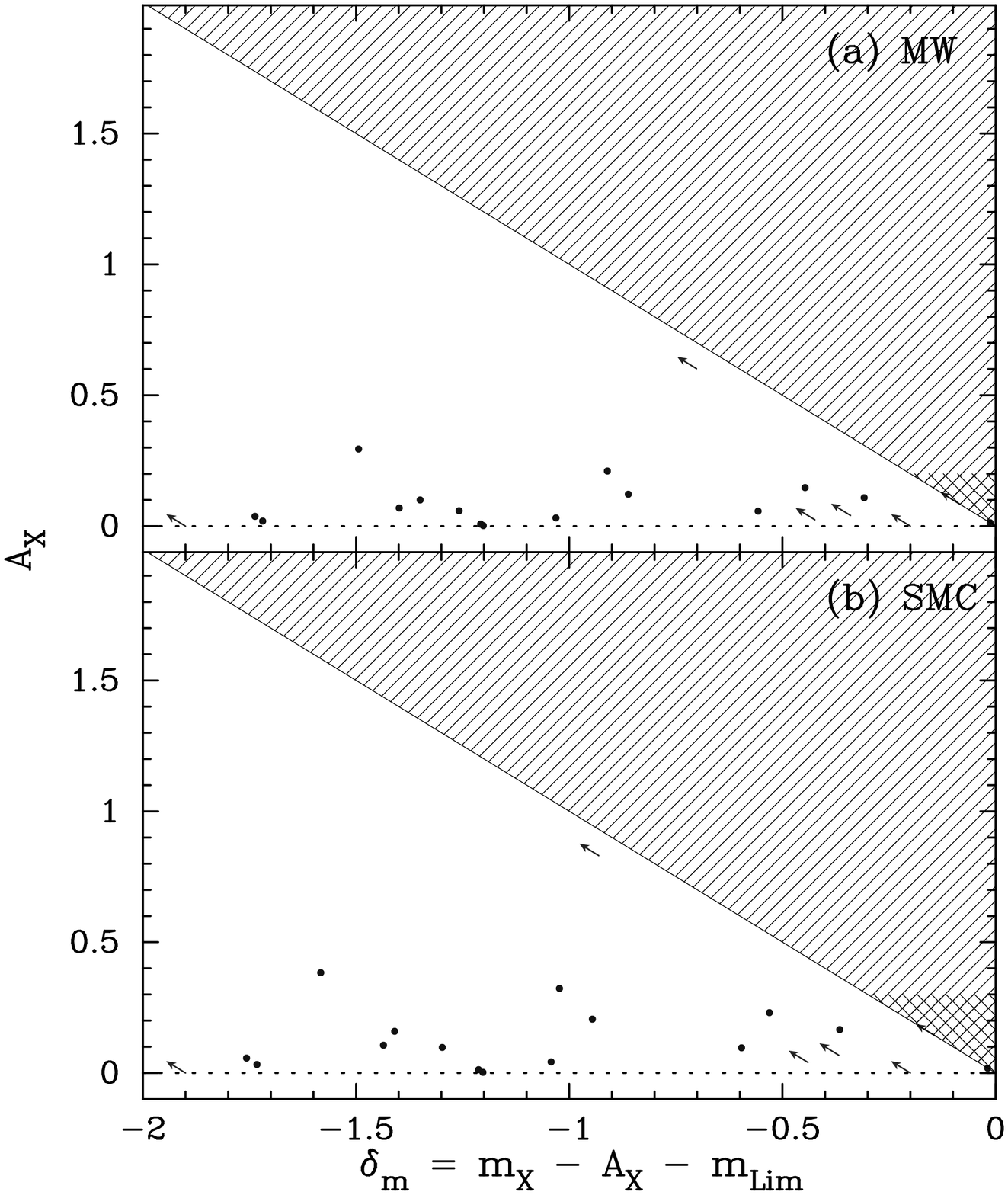}
\begin{center}
\caption{Extinction corrections $A(X)$ derived in the survey
filter at the redshift of the foreground damped \lya system
against $\delta_m$, the difference between the corrected brightness
of the background quasar relative to the limiting magnitude of
the damped \lya survey.  For the dust extinction curve, we consider
two cases: (a) the Milky Way extinction law \citep{cardelli89};
and (b) the SMC law \citep{prevot84}.  The shaded region denotes
the area of parameter space which obscured quasars would occupy.  
In the several cases where the $\N{Si^+}$ values are lower
limits, the values are plotted as diagonal arrows.
Meanwhile, the cross-hatched region designates the area of parameter space
which we contend is populated by damped \lya obscured quasars
as inferred by the observed distribution of $A(X), \delta_m$ values.
}
\label{fig:extinct}
\end{center}
\end{figure}

Figure~\ref{fig:extinct} plots the extinction corrections
$A(X)$ corresponding to the survey filter
at the redshift of the damped \lya system
(e.g. 5400\AA/[1+$z_{abs}$] for Wolfe et al.\ 1986) 
against the magnitude difference $\delta_m$ between the 
corrected magnitude of
the quasar $m(X) - A(X)$ and the limiting magnitude of the survey
$m_{lim}$.  
The shaded region denotes the parameter space where
obscured objects lie, i.e. with unobscured magnitudes
bright enough to satisfy
the magnitude limit of the survey but with observed magnitudes
obscured by the foreground damped \lya system such that $m_X > m_{lim}$.
Examining Figure~\ref{fig:extinct}, we find that the extinction values
are small and that 
the $A(X),\delta_m$ pairs fill up only a small fraction of the
allowed parameter space. 
Had the sample been significantly affected by obscuration, the distribution
of points would be continuously distributed up to the shaded region
and then cut off.  
With only one exception (the system
toward Q0458--02), all of the $A(X) < 0.5$~mag.  
Even with the SMC dust curve, the range of $A(X)$ values is
approximately $4 \times$ smaller than the spread in $\delta_m$ values.
Qualitatively, at least, we 
expect that the distribution of obscured systems follows that of
the observed distribution and therefore primarily occupies the 
cross-hatched region.  Emboldened by the results presented in 
Figure~\ref{fig:extinct}, we draw two inferences
from this exercise: (1) dust obscuration is 
biasing us against only a few damped \lya systems; and
(2) the absent systems have small $A(X)$ values, 
i.e. similar metallicity and dust depletion characteristics
as our observed sample.
It must be reemphasized that the dust-to-gas ratios we have
adopted ($\kappa^{Si}$) assumes the Si/Fe enhancements have no
nucleosynthetic contribution.  We actually expect the systems are
$\alpha$-enhanced and, therefore, the $\kappa^{Si}$ and $A(X)$ values should
be considered upper limits which further strengthens our conclusions.

Finally, consider the extinction 
for a typical sightline through the Milky Way at low redshift.
Assuming $\kappa \approx 1, \N{HI} = 10^{21} \cm{-2}$,
$R_V = 3.1$, and $\lambda_{abs} \approx 1250$\AA,
the implied extinction is 1.7~mag, i.e., more than
twice the extinction of the damped \lya system toward Q0458--02.
Dust obscuration probably plays a far more significant role at
$z<1$ than high redshift, 
particularly given the small aperture afforded by HST for 
UV spectroscopy.
Whether one must make corrections on the order of those implied
by \cite{pei95} remains to be seen, but it is obvious they must
be considered.  Maybe this explains, in part, the rarity of high
luminosity optical counterparts for the few known
low $z$ damped \lya systems \citep{turnshek01}.
Hopefully,
future studies will estimate extinction corrections 
and examine if they correlate with the optical counterparts.

\begin{figure*}[ht]
\begin{center}
\includegraphics[height=6.8in, width=5.2in,angle=-90]{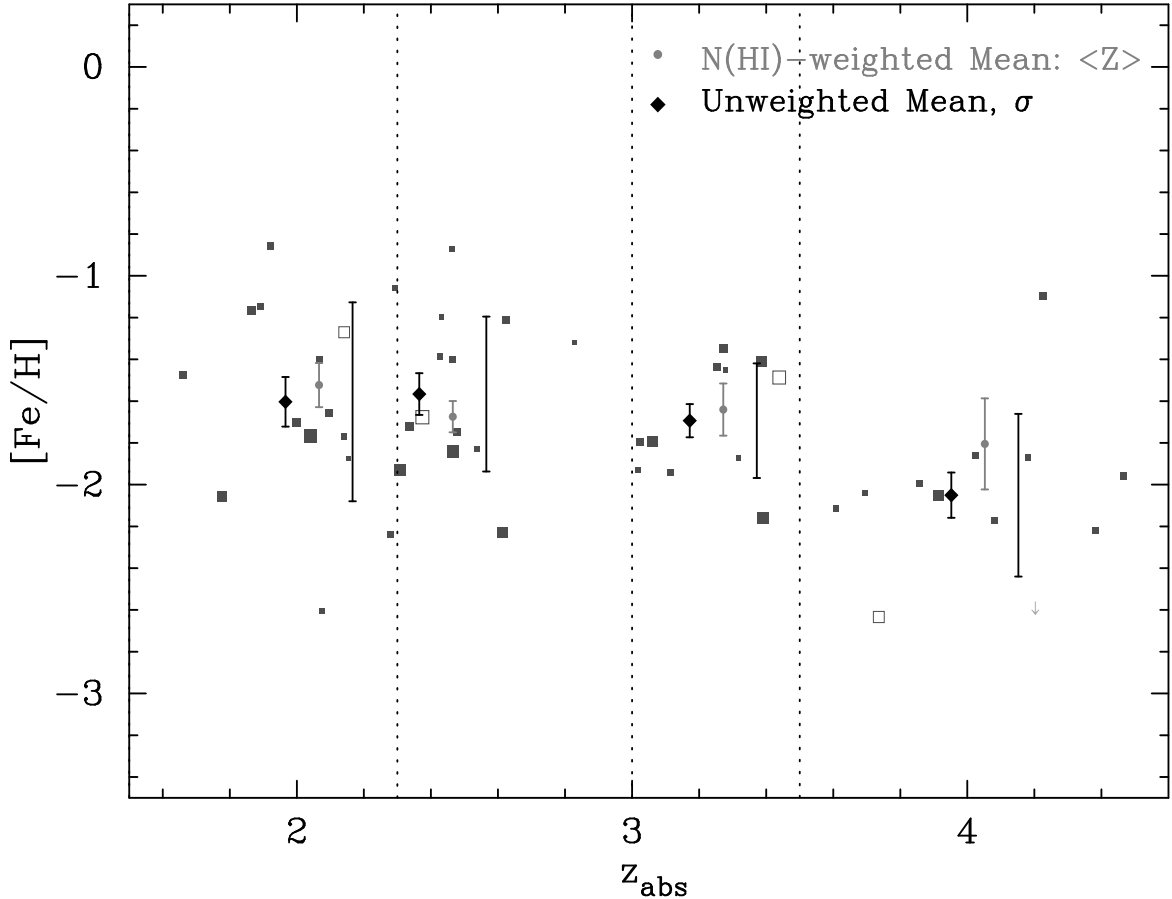}
\caption{[Fe/H] vs.\ $z_{abs}$ plot of the \nFeT\ damped \lya systems which 
comprise the entire $z > 1.7$ sample.  The open squares indicate systems
where we consider a proxy for [Fe/H] (e.g. [Ni/H]).
In all cases, the size of the data point is proportional to $\log \N{HI}$.
The dotted lines indicate 4 redshift intervals, arbitrarily chosen
to evenly divide the systems.
In each interval we plot
three statistics: (1) the HI-weighted mean metallicity $<Z>$ 
represented by the solid circle with bootstrap error.  It is plotted
at the median redshift of each interval;
(2) the unweighted mean logarithmic 
metallicity $<\feh>$ with bootstrap error depicted by the solid 
diamond; and (3) the scatter in the individual systems $\sigma(\feh)$
represented by the vertical error bar.
}
\label{fig:chemFe}
\end{center}
\end{figure*}

\subsection{[X/Zn] vs.\ [Zn/H] + $\log \N{HI}$ Revisited}

In addition to implications for dust depletion,
the results presented in Figure~\ref{fig:XZnF}  ($\S$~\ref{sec:hou})
offer insight about dust obscuration.  As argued
by \cite{hou01}, the observed anti-correlations between [X/Zn] and $\N{Zn}$
are consistent with the effects 
of dust depletion with higher depletion levels
at higher metal column densities.  
One would expect, however, 
the effects of dust obscuration to {\it minimize} the observed
trend, i.e., systems with higher $\N{Zn}$ values ought to show higher
[X/Zn] ratios (lower depletion levels) if dust obscuration were important.  
The general absence of such systems in Figure~\ref{fig:XZnF} is striking.
Perhaps the processes of 
dust formation prohibit systems with low [Zn/X] when
$\N{Zn}$ is large, but this seems contrived.
We contend this serves as further evidence that dust obscuration
is not strongly biasing the observed sample of damped systems.


\begin{figure*}[ht]
\begin{center}
\includegraphics[height=3.6in, width=2.8in,angle=-90]{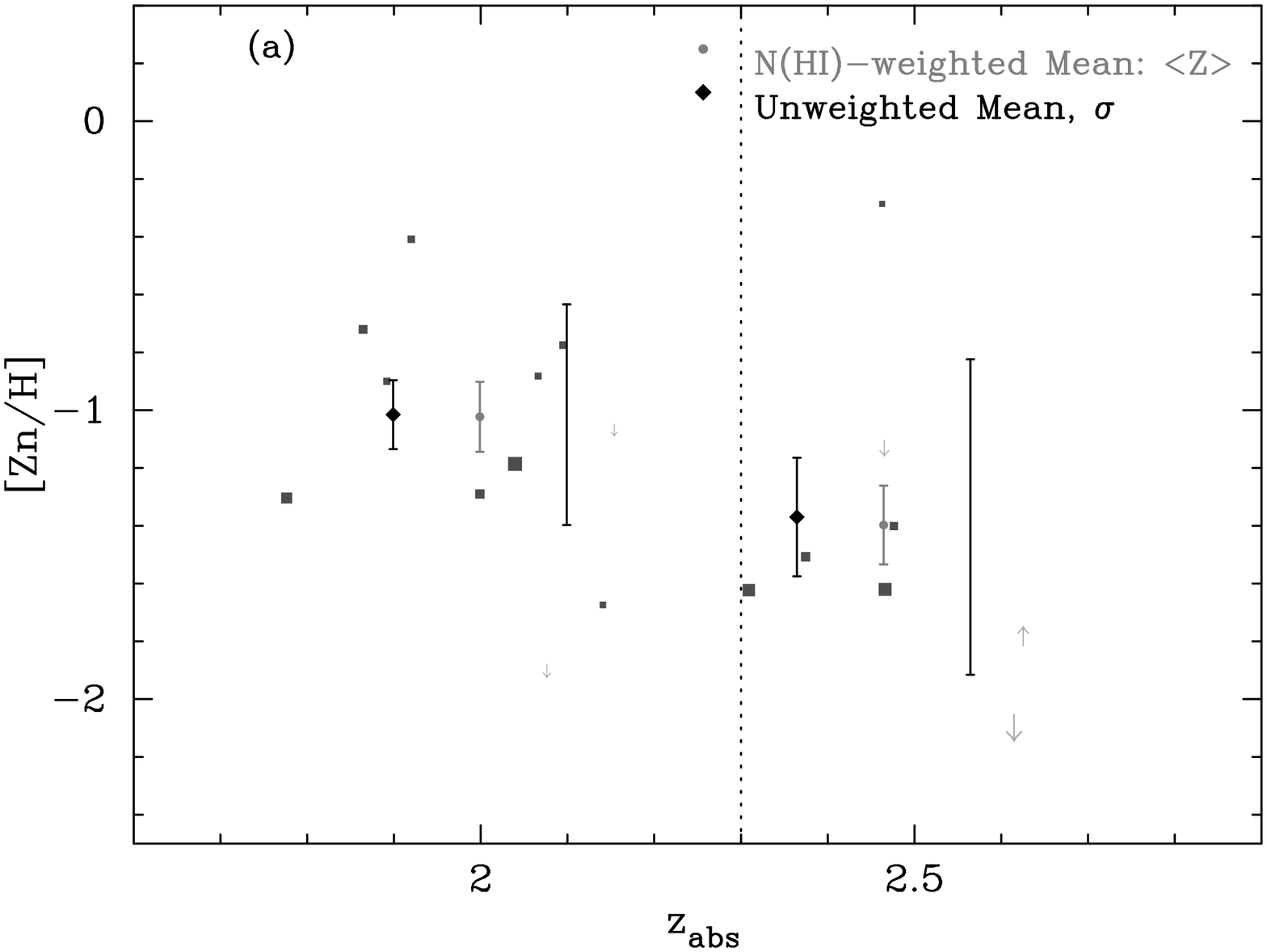}
\includegraphics[height=3.6in, width=2.8in,angle=-90]{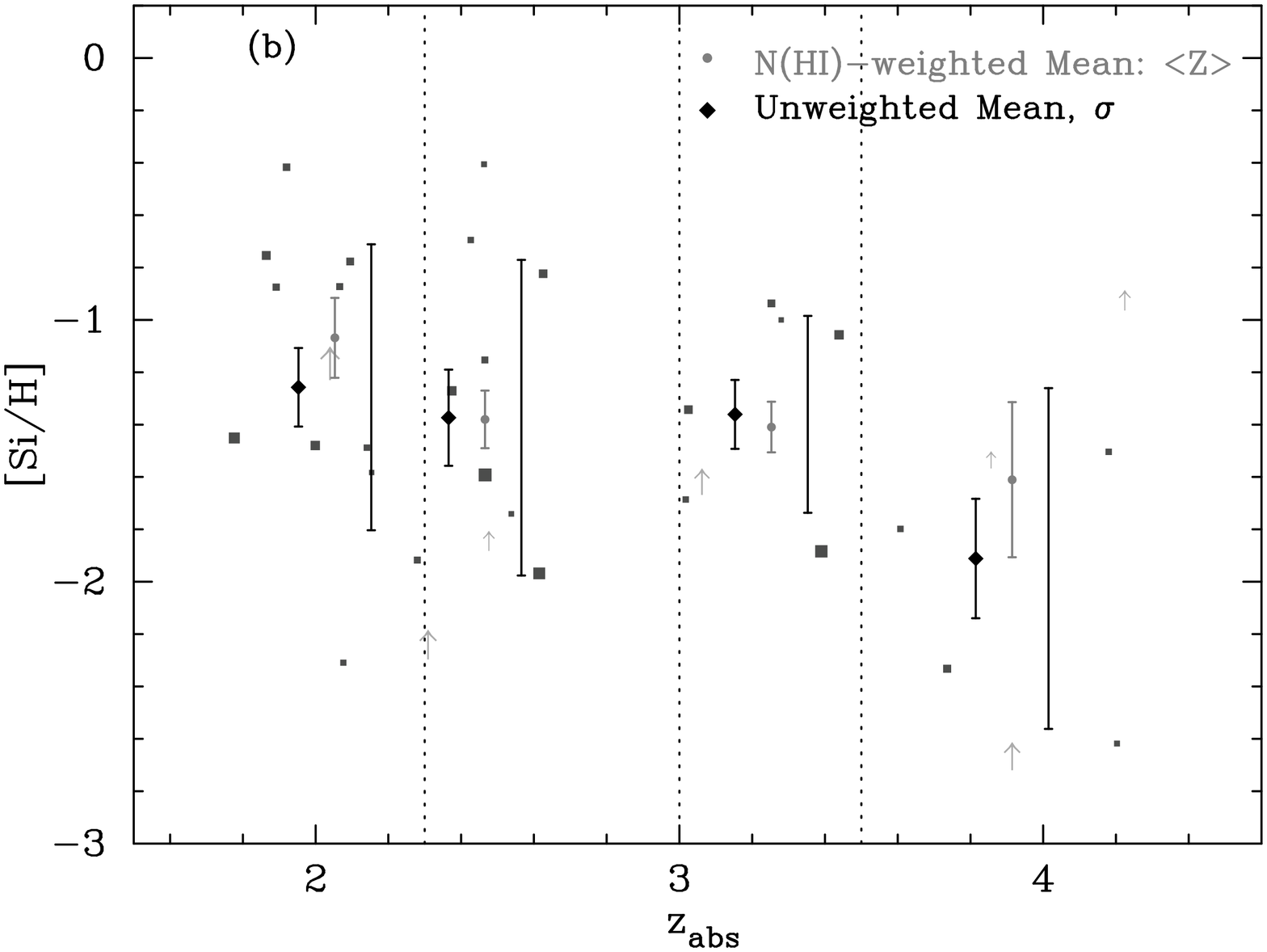}
\caption{Similar to the previous figure except we have replaced
Fe with (a) Zn and (b) Si, and we have restricted the analysis
to our HIRES sample.  In both cases, we have included limits in
the statistical analysis with their measured value, a practice which has
minimal impact on our conclusions.
}
\label{fig:chemZnSi}
\end{center}
\end{figure*}

\section{CHEMICAL EVOLUTION}
\label{sec:chemev}

In previous papers we have investigated the chemical enrichment history
of the universe in neutral gas through observations of the 
damped \lya systems (PW99; PW00; PGW01).  Although our
current efforts focus on improving the statistics at $z_{abs}>3$ with spectra
obtained from the Echellette Spectrograph and Imager \citep[ESI;][]{sheinis01},
the new HIRES sample significantly bolsters the measurements at high
redshift.  
%
Figure~\ref{fig:chemFe} presents the [Fe/H] measurements from our complete
sample including all of the systems presented in PGW01.  
Where we have adopted [Ni/H], [Al/H], or [Cr/H] as a proxy\footnote{As
discussed in $\S$~\ref{subsec:Fe}, we offset the [Ni/H], [Cr/H], and [Al/H]
values by $-0.1$, $-0.2$, and $0.0$~dex respectively.} for [Fe/H],
we have plotted the value as an open square and have assumed an additional
0.1~dex error.  
In all cases, the size of the data point is linear
with $\log \N{HI}$.  
Following our previous analysis, we focus on three statistics of the
[Fe/H], $z_{abs}$ pairs: (i) the unweighted mean of the [Fe/H] values
$<\feh>$, (ii) the $\N{HI}$-weighted mean $\ZFe$, and (iii) 
the rms dispersion $\sigma(\feh)$. 
Table~\ref{tab:chemE} summarizes these 
statistics for four redshift intervals chosen to 
evenly split the sample as well as accommodate a natural division in
the sample at $z_{abs} \approx 3$:
$I_1 \to z_{abs} \, \epsilon \, (1.5, 2.3]$, 
$I_2 \to z_{abs} \,\epsilon \, (2.3, 3.0]$, 
$I_3 \to z_{abs} \, \epsilon \, (3.0, 3.5]$, 
and $I_4 \to z_{abs}\, \epsilon \, (3.5, 4.5]$.
In Figure~\ref{fig:chemFe}, the solid diamonds correspond to $<\feh>$,
the solid circles demarcate $\ZFe$
and are centered at the HI-weighted mean absorption redshift of each
interval, and the unmarked error bars are $\sigma(\feh)$. 
For the mean statistics, the error on the points indicates the value from a
bootstrap error analysis (e.g.\ PW00).  

The results in Figure~\ref{fig:chemFe} are in good agreement with those presented
in PGW01 but with an important difference.
Previously, 
we noted a steady decrease in $<\feh>$ with
increasing redshift while $\ZFe$ remained nearly constant.  
The enlarged sample indicates a different
picture:  the $\ZFe$ and $<\feh>$ statistics are in excellent agreement and
exhibit very little evolution over the $I_1, I_2$ and $I_3$ intervals
with the $<\feh>$ statistic dropping sharply in $I_4$.
Although the weighted mean
remains nearly constant through all four intervals,
the single system toward PSS1443+27 (the data point 
with [Fe/H]~$\approx -1, z_{abs} \approx 4.2$) is heavily 
impacting both $\ZFe$ and $\sigma(\feh)$ in $I_4$.
Eliminating it -- a practice with no physical basis -- would markedly revise
these statistics because of its relatively large $\N{HI}$ and [Fe/H] 
values: $\ZFe \, = -2.07$, $\sigma(\feh) = 0.26$ for $I_4$.
In this case the $<\feh>$ statistics would match the $\ZFe$ values
at every epoch.  
Unless this is a coincidence, it suggests dust obscuration may not be
significantly affecting the statistics.

\begin{figure}[ht]
\includegraphics[height=3.8in, width=2.8in,angle=-90]{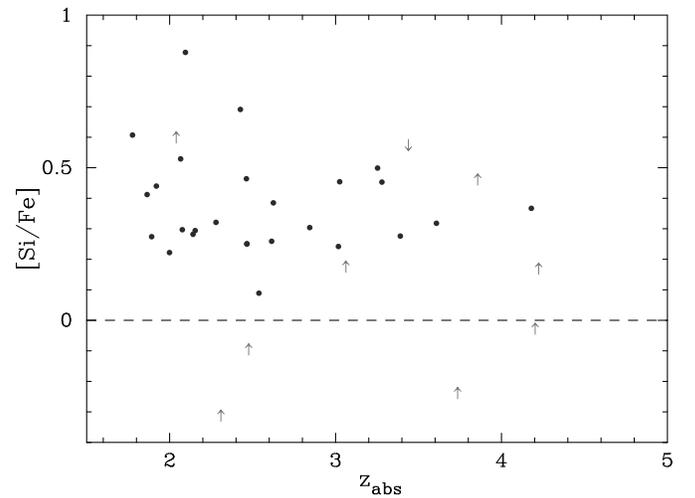}
\begin{center}
\caption{[Si/Fe] vs.\ $z_{abs}$ for our complete sample.  Although
there is significant scatter, the [Si/Fe] enhancements remain nearly
constant with redshift.
}
\label{fig:SiFevsz}
\end{center}
\end{figure}

With our HIRES observations,
one can also examine the chemical evolution of the damped \lya systems in
elements other than the Fe-peak tracers. 
In the process, one can minimize the effects of dust depletion,
but generally at the expense of introducing
uncertainties associated with nucleosynthesis.
In Figure~\ref{fig:chemZnSi} we plot
(a) [Zn/H] and (b) [Si/H] vs. $z_{abs}$ for our HIRES sample and 
Table~\ref{tab:chemE} presents the same statistics that we measured
for Fe.
The results for Zn suggest a decline in $<Z>^{Zn}$ between the $I_1$
and $I_2$ intervals but our results are dominated by small number
statistics.  The larger sample presented by \cite{ptt99} shows substantially
less evolution.  The results for Si, in contrast, are complicated by the
significant number of lower limits on [Si/H].  
For example, the sharp drop in $\ZSi$ from 
$I_1$ to $I_2$ would be lessened if the lower limits in $I_2$
were resolved.  Nevertheless, the [Si/H], $z_{abs}$ values follow similar
trends as those for [Fe/H]: a nearly constant unweighted mean from
$I_1$ to $I_3$ followed by a sharp decline at $z_{abs} > 3.5$,
a decreasing scatter (ignoring PSS1443+27), 
and a generally unevolving $\N{HI}$-weighted mean.
At present, there is more significant evolution in $\ZSi$
than $\ZFe$ which we suggest is the result of small number statistics.
We observe no evolution in 
[Si/Fe] with redshift (Figure~\ref{fig:SiFevsz})
or any other quantity and therefore expect 
equal evolution in $\ZSi$ and $\ZFe$ with complete statistics.


\begin{table*}[ht]\footnotesize
\begin{center}
\caption{ {\sc CHEMICAL EVOLUTION\label{tab:chemE}}}
\begin{tabular}{lccccc}
\tableline
\tableline
Elm&Stat & $I_1$&$I_2$&$I_3$ & $I_4$ \\
\tableline
Fe \\
& $z_{min}$ & 1.60&2.30&3.00&3.50\\  
& $z_{max}$ & 2.30&3.00&3.50&4.50\\  
& $N$ &  15& 13& 11& 12\\  
& $z_{median}$ & 2.07&2.47&3.27&4.05\\  
& $<$Z$>^{Fe}$ &$-1.523 \pm 0.106 $&$-1.674 \pm 0.075 $&$-1.641 \pm 0.125 $&$-1.805 \pm 0.218 $\\  
& $<\feh>$ &$-1.604 \pm 0.119 $&$-1.566 \pm 0.100 $&$-1.694 \pm 0.079 $&$-2.051 \pm 0.108 $\\  
& $\sigma(\feh)$ &$   0.476$&$ 0.371$&$ 0.274$&$ 0.390$\\  
\\
Si \\
& $z_{min}$ & 1.60&2.30&3.00&3.50\\  
& $z_{max}$ & 2.30&3.00&3.50&4.50\\  
& $N$ &  12& 10&  7&  7\\  
& $z_{median}$ & 2.05&2.47&3.25&3.91\\  
& $<$Z$>^{Si}$ &$-1.069 \pm 0.152 $&$-1.380 \pm 0.110 $&$-1.409 \pm 0.097 $&$-1.610 \pm 0.296 $\\  
& $<\sih>$ &$-1.258 \pm 0.150 $&$-1.374 \pm 0.184 $&$-1.361 \pm 0.132 $&$-1.911 \pm 0.228 $\\  
& $\sigma(\sih)$ &$   0.546$&$ 0.603$&$ 0.376$&$ 0.650$\\  
\\
Zn \\
& $z_{min}$ & 1.60&2.30\\  
& $z_{max}$ & 2.30&2.80\\  
& $N$ &   9&  6\\  
& $z_{median}$ & 2.00&2.46\\  
& $<$Z$>^{Zn}$ &$-1.023 \pm 0.121 $&$-1.397 \pm 0.136 $\\  
& $<\znh>$ &$-1.016 \pm 0.119 $&$-1.370 \pm 0.205 $\\  
& $\sigma(\znh)$ &$   0.382$&$ 0.546$\\  
\tableline
\end{tabular}
\end{center}
\end{table*}


\section{DISCUSSION AND SPECULATIONS}
\label{sec:discuss}

Each of the results described in the preceding sections has impact on
an important
area of damped \lya chemical abundance research.  Synthesizing
these results, we find one characteristic stands out: the {\it uniformity} of
the relative abundances of the damped \lya systems.
The observed abundance patterns -- independent of the effects of 
nucleosynthetic enrichment, dust depletion, or photoionization --
exhibit a remarkable similarity.
Furthermore, the patterns are constant not only from system to
system but also within a given damped \lya system.
In this section we discuss this homogeneity and its
profound implications for the 
protogalactic population traced by the damped \lya systems.
We conclude with some speculations on the disparity between stellar
and damped \lya abundances and offer a list of 
immediate objectives for future research.

\subsection{Uniformity of the Observed Chemical Abundances}

With few exceptions, Figures~\ref{fig:CrFe}--\ref{fig:AlFe} demonstrate
minimal dependence of X/Fe on [Fe/H]
metallicity.  Aside from Si/Fe (and Si/Cr to lesser extent),
the same conclusion holds if one considers Si/X against [Si/H]. 
Furthermore, there is no significant variation in the 
chemical abundances with redshift or HI column density. 
In short, the abundance patterns of the damped \lya systems are
strikingly similar. 
Of course, the relative abundances are not identical from
system to system; there is a statistically significant dispersion
in several of the abundance ratios.
These departures from uniformity
are likely explained by dust depletion.

By now, most of the community has concluded that 
the damped \lya systems suffer from dust depletion.  
We have been more skeptical of this conclusion
because the only evidence for dust depletion was the 
observed Zn/Fe enhancements 
and a tentative dust reddening result \citep{pei91}, the former of
which is at least partially explained by nucleosynthesis.
We are now convinced, however, that dust depletion explains
the observed [Si/Fe] vs.\ [Zn/Fe] and [Si/Ti] vs.\ [Zn/Fe] correlations 
(Figures~\ref{fig:XFe}, \ref{fig:SiTi})
as well as the [X/Si] vs. $\N{Si}$ anti-correlations presented in
Figure~\ref{fig:XSiF}.
Furthermore, the increase in [Si/Fe] at [Si/H]~$> -1$ 
(Figure~\ref{fig:XYSi}a)
is naturally accommodated within a dust depletion scenario where
higher metallicity gas has higher depletion levels.
In turn, we expect that dust depletion explains 
most of the dispersion in the Zn/Fe and Si/Fe ratios and the small
differences in the abundance patterns from system to system.
Interestingly, 
the characteristics of depletion inferred from the observations
also follow this theme of uniformity.  
First, the observed depletion levels are small ($< 0.6$~dex).  
Even in the extreme case that one 
interprets the Zn/Fe enhancements entirely in terms of depletion,
the levels are far more 
representative of Galactic halo gas or the LMC than typical
sightlines through our Galaxy.  This presumably reflects
the lower metallicity of the damped \lya systems but may
also correspond to lower density gas or physical
processes which destroy dust grains\footnote{We stress that dust
obscuration would have very little impact on the observed depletion levels.
The difference in dust-to-gas
ratio between gas with [Zn/Fe]~$= -0.5$ and [Zn/Fe]~$= -1$
is nearly negligible (Equation~\ref{eqn:kappazn}).}.
The depletion levels implied by Zn/Fe or Si/Fe do not correlate with 
$\N{HI}$ or $z_{abs}$ but only with metallicity at [Si/H]~$> -1$.  
In fact, over $60\%$ of the systems exhibit [Si/Fe]~$= 0.3 \pm 0.1$~dex
which we interpret as a floor imposed by nucleosynthesis.
Finally, there is uniformity to the dust
depletion levels within each damped system (PW96).
With very few exceptions (notably the systems toward
Q0347--38 and Q1759+75), the Si$^+$ profiles follow the 
Fe$^+$ profiles which track the Zn$^+$ profiles.  On a velocity
component by component basis, therefore, the depletion levels
are constant throughout each system.
As we argued in PW96, density fluctuations or regions shocked by
supernovae ought to imply variations in the depletion
levels.  Instead, the 
damped \lya systems tend to only show differences from galaxy to galaxy.

Because dust depletion modifies the gas-phase abundances
of the damped \lya systems, it complicates attempts to identify
the underlying nucleosynthetic patterns.
Presently, we feel dust corrections cannot be applied
to the observed abundance ratios in a quantitative
fashion.  There are too many uncertainties associated with
the composition of the dust grains, the processes of grain formation,
and even identifying the base
level of depletion.  Regarding the latter point,
previous authors have simply assumed the observed
Zn/Fe enhancements are entirely the product of dust depletion,
a presumption likely to be incorrect.
Until significant advances are made in our understanding of 
the processes of 
dust depletion, introducing dust corrections will have limited value.
Nevertheless, because
the relative abundances show little variation with metallicity or
any other property and the ratios with the largest dispersion
(Zn/Fe, Si/Fe) are the most affected by dust depletion,
this implies the nucleosynthetic patterns are very uniform.
To assess the likelihood of nucleosynthetic variations, one can examine
the Si/Fe ratios where
offsets from the solar ratio could be expected from both nucleosynthesis
and dust depletion (and even photoionization).  
Figure~\ref{fig:SiFe} indicates the observed
scatter in [Si/Fe] is small: $\sigma($[Si/Fe]) = 0.14~dex.
Because it is unlikely that dust depletion reduces the
intrinsic scatter, the nucleosynthetic
variation in Si/Fe from system to system
is probably even smaller than the observed dispersion.
This point is emphasized by the behavior of [Si/Fe] at
[Si/H]~$< -1.5$~dex (Figure~\ref{fig:XYSi}a).  At these metallicities,
essentially every system exhibits [Si/Fe]~$\approx +0.3$;
the dispersion is less than 0.08~dex and is consistent with 
measurement error.  The [Si/Fe] ratios
indicate a very uniform enrichment history for the damped \lya
systems, i.e., proportionally similar enrichment from Type~Ia and
Type~II SN.  
Allowing for the effects of dust depletion at [Si/H]~$> -1$,
this uniform enrichment history holds from [Si/H]~$-2.5 \to -0.5$,
a stunning range in metallicity.
Furthermore, the odd-even effect, as probed by Si/Al, also
shows no significant
variation from system to system.
Allowing for mild depletion in the damped \lya systems,
then, the nucleosynthetic patterns of the damped \lya
systems are essentially constant with metallicity, redshift, and $\N{HI}$.

The principal result of our examination of dust obscuration is
that the majority of damped \lya systems have very similar 
extinction properties; all but one system has an 
estimated extinction correction $A(X) < 0.5$~mag.  
Although this one outlier demonstrates that 
systems with significant
extinction exist, the majority of the observed sample -- and by
inference we contend
the majority of all damped \lya systems -- have minimal extinction.
This implies that systems hidden from the damped \lya surveys
have similar characteristics to our observed sample, i.e., 
the observed uniformity in chemical abundances is not the result
of dust obscuration.

Finally, we stress that the mean metallicity of the damped \lya systems
exhibits a temporal constancy.  Both
the unweighted and $\N{HI}$-weighted mean metallicities
demonstrate only mild evolution from $z = 1.5 \to 3.5$, an epoch
spanning $\approx 2$~Gyr where significant star formation is known to occur
\citep{steidel96}.  There appears to be a marked decline
at $z>3.5$, but by only the factor of a few.  There
may also be a decline in the dispersion of [Fe/H] values
with increasing redshift, but this difference is also small.  With
the possible exception of the very highest redshift systems, therefore,
the damped \lya systems have an unevolving chemical enrichment history.

\subsection{Interpreting the Uniform Abundance Patterns}

The uniformity of the chemical abundances of the damped \lya
systems has profound implications for their 
enrichment history.  It must be emphasized 
that the damped \lya systems are {\it selected in absorption
only according to their HI column density}.  With the exception of dust
obscuration -- an issue we feel increasingly confident has minimal
impact at high $z$ -- this selection criterion is unbiased.
In turn, one expects the damped \lya
systems to correspond to the bulk of the protogalactic population,
i.e., galaxies with a large range of mass, age, morphology,
and chemical enrichment history.  
The actual distribution of protogalaxies is determined by the
product of their number density and HI cross-sectional area 
(i.e. their optical depth, $\tau \equiv n \sigma_{HI}$). 
The observed chemical abundances, meanwhile, not only depend on
the base properties of the galaxies, but
also on gradients and variations within each galaxy.  For example,
sightlines will preferentially pass through the outer
regions of galaxies with high central HI surface density simply
because the available cross-section increases radially as $r^2$ but are
restricted to the inner regions of systems with low central HI
surface density.
Therefore, one has every reason to expect large variations in the
chemical abundances of the damped \lya systems from system
to system and within a given system.  

In terms of the total
metallicity of individual damped \lya systems, a significant dispersion
is observed 
(e.g. Figure~\ref{fig:chemFe}).  It is most pronounced at $z<3$ which
probably reflects protogalaxies with a greater range of mass and age
than at $z>3$.
It is also more pronounced for sightlines with 
$\N{HI} < 10^{21} \cm{-2}$ which might indicate metallicity gradients
within a given system. 
In terms of the relative abundances, however, the
absence of significant variation is striking. 
At most, there are differences of $\approx 0.3$~dex and these are 
almost entirely the product of dust depletion.  This uniformity 
raises the following questions:
Granted the dispersion in [Fe/H] and $\N{HI}$ values, why are
the chemical abundance patterns and therefore the dust depletion
and nucleosynthetic properties so similar?  Furthermore, why do
these properties exhibit such minimal evolution with redshift?

One possibility is that the damped \lya systems are dominated by a
subset of protogalaxies which have similar physical characteristics.
Perhaps they arise in galaxies with the largest HI radial
extent \citep[e.g.][]{mmw98} or in a very large number of low mass galaxies.
In this scenario, the range of $\N{HI}$ and [Fe/H] values 
would result from gradients within each galaxy.
The systems would have similar enrichment histories
and therefore possess similar nucleosynthetic patterns.
It might be difficult to explain the constant
mean metallicity with redshift, but perhaps this sub-population
of galaxies has a small overall star formation rate.
A more dramatic scenario to explain the relative abundance
uniformity is that the nucleosynthetic histories of {\it all}
protogalaxies at [Fe/H]~$< -1$ are the same.
The protogalactic abundance patterns we observe are derived
from the convolution of the initial mass function (IMF) and the
integrated star formation history.  
Although the concept of a universal IMF is widely supported,
one might expect significant differences in the level 
and duration of star formation throughout the entire galactic 
population.  In fact, it is these variations which presumably generate
the differences in the observed abundances of various stellar
populations.   In order to explain the uniformity of the abundance
patterns in this scenario, therefore, one may require
a disparity between the damped \lya
and stellar abundances which we consider in the following
sub-section.

While the constancy of the chemical abundances from system to
system has important implications for the properties of high
redshift protogalaxies, 
the uniformity within each system (as a function of velocity along 
the sightline) may place the tightest constraints on their physical nature.
As first emphasized in PW96, the dominant ionic species (e.g. Fe$^+$, Si$^+$)
show no significant variations from velocity component to
component.  This uniformity is not universal -- the system toward
GB1759+75 is an excellent counter-example \citep{pro02} --  but holds
for nearly every damped \lya system in our sample.
Therefore, there is little variation 
both in terms of dust depletion and nucleosynthesis
at the presumably kpc scales resolved by the velocity profiles.  
If the observed variations
in $\N{HI}$ and [Fe/H] are at least partly due to gradients
in surface density and enrichment in these protogalaxies,
then there must be very little radial variation in dust depletion
or nucleosynthetic enrichment.  Under the assumption that the
absorption arises within a single protogalaxy, the observations 
indicate one or more of the following: 
(i) The mixing timescales are sufficiently small to yield constant
relative abundances in these protogalaxies.  For this to hold true,
the timescale would have to be shorter than both the timescales
for supernovae with significantly different yields (e.g. $\lesssim 1$~Gyr
for Type~Ia SN) and the timescale for dust grain formation;
(ii) Abundance variations are washed out in velocity space
because the observed velocity components are actually the 
superposition of many 'clouds', each with its own intrinsic 
abundances and dust depletion.  Of course, it could be difficult
to explain why the mean abundances of each component are so similar;
(iii) The protogalactic volume probed by a damped \lya observation 
is smaller than the scale where abundance differences are large.
This scenario would require that the chemical filling-factor of
protogalaxies to be very small at these epochs, an unlikely presumption.
Furthermore, it would be very difficult to explain the 
kinematic characteristics observed from the low-ion profiles \citep{pro97b}.
Of these three possibilities,
we believe efficient mixing is the most likely explanation.

On the other hand, if the damped \lya profiles result from the superposition
of absorption from several protogalaxies, then the observed
uniformity places important constraints on the properties of the
various 'clumps'. 
The most important result from our first
study on damped \lya velocity fields
\citep{pro97b} is that protogalaxies as single disks within
the CDM framework cannot reproduce our observations.
This conclusion contradicted the protogalactic paradigm of the time
\citep[e.g.][]{kau96,mmw98} and inspired alternate protogalactic models
within the CDM cosmology \citep{hae98,maller01}.  The key
difference of the new scenarios is that the velocity profiles of
the damped \lya systems arise from the intersection with several 
protogalaxies.
If these protogalactic clumps have formed
independently, they should have unique metallicities, enrichment
histories, and depletion levels.  
It may be very difficult to account for the chemical
similarity among the low-ion
profiles within this scenario.  For the metal-poor systems 
(i.e. [Si/H]~$<-1.5$), where we note little dispersion in any abundance
ratio from sightline to sightline, the multiple clump scenario may
not be challenged.  The systems at [Si/H]~$>-1.5$, in contrast,
do exhibit variations in several ratios (e.g. Si/Fe and Zn/Cr) 
from system to system and
might therefore be expected to show variations along the velocity profile.
We emphasize that these systems tend to have larger velocity 
widths \citep{wol98}, in fact these are the systems which {\it require}
the multiple clump scenario within the CDM framework.
At present, we cannot test the multiple clump scenario against our
observations because these models provide at best crude predications
for the dust properties and enrichment histories of the protogalactic
clumps.  We suspect, however, that these observations may ultimately
pose a great challenge for theories of galaxy formation within the
CDM cosmology.

\subsection{Gas-phase/Stellar Abundance Disparity}

The damped \lya systems are the dominant neutral gas 
reservoirs of the universe.
Therefore, it is natural to identify them as the fuel for past and
present star formation and compare their abundance patterns with
current stellar populations. 
An important property of the damped \lya systems, however, is the
fact that the majority have a very small molecular fraction \citep{petit00}.
If the observed gas in the damped \lya systems does fuel
star formation, then there must be an intermediate stage which inspires
the growth of molecular clouds.  
This suggests star formation is disconnected from the ISM
of protogalaxies as probed by the damped \lya systems.
This disparity may manifest itself through contradictions in the
abundance patterns of the damped \lya systems with present-day
stellar populations.  For example, although we contend
the damped \lya systems
exhibit $\alpha$-enhancements similar to the Galactic
halo stars, the damped \lya Si/Al ratios are substantially lower
than the observed stellar values.
Furthermore, we have identified variations along the Fe-peak 
which may contradict the chemical abundance patterns of 
every stellar population.

Our observations might indicate that star 
formation occurs independent of the gas probed by our observations.
This is not a profound supposition; star formation in
the Milky Way is primarily restricted to dense molecular clouds whose
filling factor is small.  In this case, enrichment occurs independent
of the global ISM and can proceed on very short timescales such that 
the stellar by-products have unique abundance patterns.
We envision a scenario of star formation where 
the first stars to form in a molecular cloud should have abundances
similar to the ISM, but star formation proceeds rapidly through
several generations with the subsequent
stellar populations more closely reflecting the rapidly changing
abundances of the molecular cloud.  Although supernovae feedback
will ultimately distribute the metals of the molecular cloud to the
ambient ISM, the effects on the mean abundances might be small.
This would be particularly true if the majority of metals are
carried out to the local IGM instead 
\citep[e.g.][]{maclow99,kubulnicky97}.
In short, one can envision an ISM mildly enriched by the by-products
of molecular clouds, each with its own metallicity and relative 
chemical abundances giving mean abundances different from 
observed stellar populations.

Star formation may also have significant impact on the protogalactic
population selected by damped \lya systems.
There are two competing processes affecting the damped \lya 
cross-section of a given galaxy.
As galaxies form through the accretion and dissipation of
gas, their HI surface densities will increase until they have
an appreciable surface density with $\N{HI} > 2 \sci{20} \cm{-2}$.
On the other hand, star formation lowers the surface density
through the consumption of gas and the feedback from supernovae.
Qualitatively, we expect the damped \lya systems to more 
frequently probe galaxies in the early stages of formation, in
particular prior to significant star formation.  This might well
explain the lack of chemical evolution observed in the damped 
systems: at every epoch the damped \lya systems are predominantly
gas-rich, metal-poor galaxies.
Systems which have experienced bursts of
star formation (e.g. the Lyman break galaxies) may have consumed
or blown out an appreciable fraction of their HI gas and have a 
comparatively small damped \lya cross-section.


\subsection{Future Directions/Objectives}

In closing, we present an abundance wish list for
the damped \lya systems.  
Achieving these goals will markedly improve our understanding
of the chemical abundances of the damped \lya systems.  We expect
most, if not all, will be accomplished in the next five years 
through the active efforts of our group and others throughout the
community.

\begin{enumerate}

\item A comparison of the relative abundance of the $\alpha$-elements
(Si, S, O, Ar, Ti) in several damped \lya systems.  In this paper,
we have already demonstrated the importance of Si/Ti and S/Si
for considerations
of dust depletion.  Comparisons among the $\alpha$-elements will
provide further insight into photoionization, nucleosynthesis, and dust
depletion.

\item Measurements of Mn at $z_{abs} > 1.5$.  We must ascertain
if Mn/Fe is universally sub-solar in the damped \lya systems.

\item Relative abundance measurements for a sample 
of $z_{abs} < 1$ damped systems at $\lambda_{rest} < 1500$\AA.  
Although this will require space-based observations, these measurements
are critical to unveiling the evolution of galactic chemical abundances.

\item An investigation of the odd-even effect at all metallicities.
This will require a number of P~II measurements in the more metal-rich
systems.

\item  Empirical evidence for/against photoionization.  With the
advent of UV-sensitive ($\lambda < 4000$\AA)
echelle spectrographs, one should be able
to examine ions like Fe~III and C~III to constrain the effects
of photoionization.

\item  Over 100 damped \lya metallicity measurements.  These
observations are still the most efficient means for testing
chemical evolution scenarios at any epoch. 

\item  A survey of radio-selected quasars for damped \lya systems.
This is the ideal approach for assessing the impact of dust obscuration
on the damped \lya surveys.
\cite{ellison01b} have been pursuing such a program and the first
results are forthcoming.

\item  Better oscillator strengths for a number of transitions
including Fe~II 1611. 

\item  A measurement of at least one element with Z$> 30$.  This will
require a very special damped \lya system, i.e., one with very high
HI column density and metallicity in front of a very bright quasar.

\item Additional emphasis on the systems with [Si/Fe]~$> +0.4$~dex to
better evaluate the dust characteristics of the damped \lya systems.

\end{enumerate}

\acknowledgments

The authors wish to extend special thanks to those of Hawaiian ancestry 
on whose sacred mountain we are privileged to be guests.  Without 
their generous hospitality, none of the observations presented 
herein would have been possible.
We thank Chris Howk, Max Pettini, and Eric Gawiser
for stimulating discussions and very helpful comments.
J.X.P. acknowledges support from a
Carnegie postdoctoral fellowship. AMW was partially supported by 
NSF grant AST 0071257.

\appendix

\section{Dust Corrections}

In this section we examine the implications of the dust correction
formalism introduced by \cite{vladilo98}.  The principle premise
of his treatment is that dust in the damped \lya systems has the
same average number of atoms of each element per grain as Galactic
dust.  As we show below, this has the effect of minimizing departures
from solar relative abundance for our observed abundance ratios.

Let us represent the number of atoms of Fe in the gas and dust phases
as $n^{Fe}_{g}$ and $n^{Fe}_{d}$ respectively.  For a system that is
90$\%$ depleted, this implies the dust fraction 
$\fxi{Fe} \equiv n^{Fe}_d / (n^{Fe}_d + n^{Fe}_g) = 0.9$.
Now consider the abundances of an element X with a dust fraction
$\fxi{X} \equiv n^X_d / (n^X_d + n^X_g)$.  Assume that in the Galactic
ISM element X depletes relative to Fe like

\begin{equation}
\theta^X = \frac{n^X_d}{n^{Fe}_d} \perd
\end{equation}
The prescription introduced by \citep{vladilo98} is to 
assume that this ratio applies to the damped \lya systems
{\it independent of the intrinsic abundances of the system}.  
This has important implications for the depletion of element X
relative to Fe in the damped \lya systems as inferred from their
gas-phase abundances.
Consider the relative dust fraction of element X to Fe:

\begin{eqnarray}
D^X &\equiv \frac{\fxi{X}}{\fxi{Fe}}\\
&= \frac{n^X_d}{n^X_d + n^X_g} \cdot \frac{n^{Fe}_d}{n^{Fe}_d + n^{Fe}_g}\\
&= \frac{\theta^X n^{Fe}_d}{\theta^X n^{Fe}_d + n^X_g} 
       \cdot \frac{n^{Fe}_d}{n^{Fe}_d + n^{Fe}_g}\\
&= \frac{1 + \frac{n^{Fe}_d}{n^{Fe}_g}}{ \frac{n^X_g}{\theta^X n^{Fe}_g}
  + \frac{n^{Fe}_d}{n^{Fe}_g}}
\end{eqnarray}
The last expression demonstrates that
$D^X > 1$ if $n^X_g / n^{Fe}_g < \theta^X$, i.e. if
the observed gas-phase ratio of X/Fe is less than 
the dust composition in the Galactic ISM.
Therefore, elements with low [X/Fe] values will tend to
have larger dust corrections which serve to minimize any departure
of X/Fe from the solar abundances.
Conversely, $D^X < 1$ if the observed ratio of X/Fe is greater than the
dust ratio in the ISM.

It is best to illustrate this effect with an example.  In the
ISM, Mn has a similar refractory nature as Fe: 
$\fxi{Mn}_{ISM} = \fxi{Fe}_{ISM}$.  This implies that 

\begin{equation}
\theta^{Mn} = \frac{n^{Mn}_d}{n^{Fe}_d} \approx \frac{10^{\e{Mn}}}{10^{\e{Fe}}}
\approx \frac{1}{100}
\end{equation}
where $\e{Mn}$ and $\e{Fe}$ are the solar abundances of Mn and Fe.
In the damped \lya systems, one observes $n^{Mn}_g / n^{Fe}_g \approx 1/200$
and therefore $D^{Mn} > 1$ implying a higher fraction of Mn is in the 
dust phase than Fe.  
Therefore, when one applies a dust correction to the observed
Mn/Fe ratio, the resulting value is closer to the solar ratio
of $\approx 1/100$.  
A similar effect holds for a mildly refractory element like Si.
Because the Si/Fe ratio observed in the damped systems is much larger
than $\theta^{Si}$ from the ISM, the dust correction to Si is much smaller
than for Fe and the observed Si/Fe is corrected downward to the solar value
when dust corrections are applied.

In short, the prescription introduced by \cite{vladilo98}, which makes
an nonphysical assumption about the formation of dust grains, tends
to minimize observed departures from solar for all of the X/Fe ratios.


\end{document}